\documentclass[10pt,journal,final,twocolumn]{IEEEtran}
 
\usepackage{multirow}

\usepackage{amsmath}
\usepackage{subfigure}
\usepackage{amsthm}
\usepackage{setspace} 
\usepackage{psfrag}
\usepackage{amsfonts}
\usepackage{amssymb}
\usepackage{stmaryrd}
\usepackage{mathrsfs}
\usepackage{epstopdf}
\usepackage{enumerate}
\usepackage{booktabs}  
\usepackage{amsmath, color}
\usepackage{dsfont,bm}
\usepackage{amssymb}
\usepackage{epsfig}
\usepackage{graphicx}
\usepackage{amssymb}
\usepackage{epsfig}
\usepackage{epstopdf}
\usepackage{subfigure}
\usepackage{balance}
\usepackage{enumerate}
\usepackage{verbatim,algorithm,algorithmic,cite}
\usepackage{booktabs}
\ifCLASSINFOpdf
\else
\fi
\newtheorem{remark}{Remark}
\newtheorem{theorem}{Theorem}

\newtheorem{lemma}{Lemma}

\newtheorem{corollary}{Corollary}

\newtheorem{proposition}{Proposition}

\allowdisplaybreaks[4]

\begin{document}
	%
	\title{Relative Entropy-Based  Constant-Envelope Beamforming for Target Detection in Large-Scale MIMO Radar With Low-Resoultion ADCs}

\author{\IEEEauthorblockN{Ziyang Cheng, \IEEEmembership{Member,~IEEE}, 
Linlong Wu, \IEEEmembership{Member,~IEEE},
Bowen Wang, \IEEEmembership{Student Member,~IEEE}, \\
Julan Xie, \IEEEmembership{Member,~IEEE},
and Huiyong Li
\vspace{-0.6em}
}\\
\thanks{
Copyright (c) 2015 IEEE. Personal use of this material is permitted. 
However, permission to use this material for any other purposes must be obtained from the IEEE by sending a request to pubs-permissions@ieee.org.
}
\thanks{
The work of Z. Cheng, B. Wang, J. Xie and H. Li was supported in part by the National Natural Science Foundation of China under Grants 62001084, 62231004 and 62031007.
The work of L. Wu was supported in part by ERC AGNOSTIC under Grant EC/H2020/ERC2016ADG/742648, and in part by FNR CORE SPRINGER under Grant C18/IS/12734677.
The review of this article was coordinated by Prof. Chau Yuen.
(\emph{Corresponding author: Linlong Wu})}
\thanks{Z. Cheng, B. Wang, J. Xie and H. Li are with School of Information and Communication Engineering, University of Electronic Science and Technology of China, Chengdu 611731, China. (Email: zycheng@uestc.edu.cn, B\_W\_Wang@163.com, julanxie@uestc.edu.cn,  hyli@uestc.edu.cn).}
\thanks{L. Wu is with the Interdisciplinary Centre for Security, Reliability and Trust (SnT), University of Luxembourg, Luxembourg City L-1855, Luxembourg. (Email: linlong.wu@uni.lu).}
}
	
	\maketitle
	%



	\begin{abstract}
 	Hybrid digital/analog architecture and low-resolution analog-to-digital/digital-to-analog converters (ADCs /DACs) are  two low-cost implementations {\color{black}for large-scale} millimeter  wave (mmWave)  systems. {In this paper, we investigate the problem of {\color{black} constant-envelope transmit}  beamforming for large-scale multiple-input multiple-output (MIMO) radar system, where the transmit array adopts a hybrid digital/analog architecture with a small number of RF chains and the receive array {\color{black}adopts a fully} digital architecture with low-resolution ADCs.  We derive the relative entropy between the {\color{black}probability density functions}   associated with the two test hypotheses under low-resolution ADCs. We formulate our optimization  problem by maximizing the relative entropy, subject to the constant envelope and orthogonality constraints. {To suboptimally solve the resultant  problem,  a two-stage framework is developed.  In the first stage, we optimize  the {\color{black}transmit power} at the directions of the target and {\color{black}clutter}.  In the second stage, an efficient iterative algorithm based on {\color{black}majorization-minimization}  is presented to obtain the constant-envelope beamformer according to the attained {\color{black}transmit power}.} Specifically,  we apply a quadratic function as the minorizer, leading to a	low-complexity solution {\color{black}at  each} iteration. In addition,  to further facilitate low-cost implementation of the constant-envelope beamformer, we consider the problem of one-bit   beamforming design  and propose an  efficient {\color{black}iterative} method based on the  Nesterov-like gradient method to solve  it. Numerical simulations are provided to demonstrate the effectiveness of the proposed schemes. }
	\end{abstract}
	

	\IEEEpeerreviewmaketitle


	%
	\IEEEpeerreviewmaketitle

\section{Introduction}
\IEEEPARstart{M}{illimeter} wave (mmWave) {\color{black}technology has received extensive attention \cite{8753573,8114253,8642926,8835577}, as a  promising candidate that can settle the current challenges  of bandwidth shortage and large antenna size  for the vehicle radar  systems.}
Due to the shorter wavelength at mmWave frequencies, 
{more antennas can be placed in the same array size} {\color{black}to achieve} highly directional beamforming. This results in the large-scale concept for mmWave  systems. Nevertheless, it is impractical to adopt the conventional fully digital beamforming architecture with high-resolution analog-to-digital/digital-to-analog converters (ADCs/DACs) for the large-scale antenna array. The reason is that, in conventional fully digital beamforming,  each antenna requires one radio frequency (RF) chain with high-resolution ADC/DAC,  which leads to prohibitive cost and power consumption for mmWave  systems\cite{4977467,4561519,1367565}.  

\vspace{-1em}
\subsection{Hybrid Beamforming Structure}
To address the hardware limitation of conventional fully digital architecture, one potential way is to use the hybrid analog/digital  scheme in which  a small number of RF {\color{black}chains} is utilized to implement the digital (baseband) beamformer and a large number of phase shifters to realize the analog beamformer. This scheme has been widely considered for mmWave communication systems \cite{zhang2014achieving,6717211,liang2014low,dai2015near,yu2016alternating,han2015large,sohrabi2016hybrid}. For example,  the authors in \cite{zhang2014achieving} propose  a   hybrid precoding scheme  for a special case in which the number of RF {\color{black}chains} is larger than twice  that of data streams.  In addition,    \cite{yu2016alternating} proposes an alternating  method    to optimize the hybrid beamformer for mmWave massive multiple-input multiple-output (MIMO) communication systems.  
Recently, a hybrid beamforming architecture is also considered for radar systems with sparse array\cite{9237135}, where the transmit and receive hybrid beamformers are jointly optimized to achieve the desired {\color{black}elevation-azimuth  image}. Additionally, the learning approach is proposed in  \cite{9413904} to synthesize the probing beampattern for mmWave hybrid   beamforming based MIMO radar system. Further, the hybrid beamforming structure is applied in the integrated sensing and communications \cite{9950549,cheng2022hybrid}.

\vspace{-1em}
\subsection{Low-resolution ADC/DAC}
Another possible solution  is  the usage of low-resolution ADC/DAC  (e.g., 1-3 bits) at each antenna,  as the ADC/DAC power increases exponentially with its resolution \cite{1550190}. For instance, in \cite{7094595,6987288,8850102,8171203,7355388}, the received signal at each antenna is directly quantized by low-resolution ADCs, and the  corresponding receive digital beamforming schemes are considered without any analog beamforming. Besides, some works are devoted to investigating the quantization performance of  low-resolution ADCs for MIMO systems. The authors in \cite{4407763} propose the {\color{black}additive quantization noise model (AQNM)} to fit the low-resolution quantization error for wireless systems. Using this model, the effects of the number of ADC bits on the uplink rate when considering Nakagami-$ m $ fading channel and {Rayleigh} fading channel are analyzed in \cite{8640825}  and  \cite{7307134}, respectively.
In addition, the achievable rate and energy efficiency are analyzed based on the AQNM for mmWave communication system with hybrid and digital beamforming (DBF) receivers with low-resolution ADCs in \cite{7961157}, where the results show that  in a low SNR regime, the performance of DBF with 1-2 bit ADCs outperforms HBF. Moreover, {\color{black}the authors   in \cite{8310639}} extend the work in  \cite{7961157} to the scenario of multi-user (MU) and imperfect channel state information (CSI). More recently, the hybrid architecture with low-resolution DACs/ADCs  at both transmitters and receivers are considered for mmWave communication systems in \cite{9110851}.
	
As for radar applications  with low-resolution ADCs/DACs,  the works mainly concentrated on DoA estimation. For instance, an algorithm is proposed in \cite{2020Direction} to reconstruct  the unquantized  measurements from one-bit samples, followed by the MUSIC method to estimate the DOAs of targets. In addition, The maximum likelihood (ML)   
is developed in \cite{9112678} for finding  DoA and velocity of a target  from one-bit data. Recently,  the problem of DoA estimation from one-bit observations is investigated for a sparse linear array   in  \cite{9585542}. In addition, \cite {9780031} analyzes the theoretical detection gap between the one-bit and infinite-bit  MIMO radars and then jointly designs the waveform and filter to achieve the theoretical performance.
	
Although the above works have studied the problem of {\color{black}parameter estimation} for one-bit radar systems,    the problem of transmit design with  hybrid beamforming and low-resolution ADC architecture  for mmWave radar detection  has not yet been studied. To achieve a low cost and a low-complexity implementation for a large-scale mmWave radar system with the functionality of target detection, 
we explore a combination of  the two promising schemes, where hybrid architecture is utilized at the transmitter, while low-resolution ADCs are adopted at the receiver.

\vspace{-1em}
\subsection{Information-theoretic  Criterion for Radar Systems}
On the other hand, in radar applications,  the detection ability  can be measured {by information-theoretic quantities, including}  mutual information, relative entropy, etc. Recently, {\color{black} information-theoretic criteria have} acquired extensive
attention {\color{black}in radar  systems}. The pioneering work of
Woodward \cite{woodward2014probability} first proposes the utilization of   information theory to radar receiver
design. Later, the information-theoretic design criterion for a
single waveform by exploiting a weighted linear sum of the
mutual information  between target radar signatures and the
corresponding received beams is presented in \cite{4200705}.  
Apart from these, some interesting extensions including mutual information or relative entropy-based waveform design in the presence of clutter \cite{4200715,5467189,6025317,7086341,8101018,naghsh2017information}  emerge thereafter. 
Nevertheless, information-theoretic criteria is rarely considered for  the large-scale mmWave radar system with low-resolution ADCs. 
Particularly, the resulting objective using information-theoretic criteria with low-resolution ADCs is far more complicated.  
The resultant problem in this work is, thus, more challenging to solve than the problem with ideal ADCs.

	\vspace{-1em}
	\subsection{Contributions and Notations}
	In this paper,  we focus on transmit beamforming design via relative
	entropy maximization for large-scale mmWave MIMO system in the presence of (signal-dependent) clutter,  the transmit array adopts a hybrid digital/analog architecture with a small number of RF chains and the receive array adopts fully digital architecture with low-resolution ADCs.
	
	Specifically, the main contributions of this work are summarized as follows:
	{
		\begin{itemize}
			\item For the low-cost architecture with transmit hybrid architecture and low-resolution ADCs, we derive the relative entropy between the probability density functions (PDFs) associated with the hypotheses ${\cal H}_0 $ and  ${\cal H}_1$ based on the AQNM, and simplify the relative entropy using some asymptotic results. With the criterion of the relative entropy  maximization, we formulate the problem of  transmit  beamforming subject to the constant-envelope and orthogonality constraints. To the best of our knowledge, the problem of transmit design with the proposed novel architecture for mmWave radar detection has not yet been studied.


			\item To {handle}  the intractable optimization problem,   a  {two-stage}  optimization framework  is proposed. To be more specific, we first optimize the {\color{black}transmit power} at the directions of target and {\color{black}clutter}, and then design the constant-envelope beamformer according to the obtained {\color{black}transmit power}.  Furthermore, {in the second stage}, we propose  a quadratic function  to minorize the quartic objective function
			based on the minorization-maximization (MM) framework \cite{7547360,6601713,wu2017transmit}, and obtain  a closed-form
			solution for the quadratic programming problem at every
			iteration. 
			
			\item To  further facilitate low-cost implementation of the system, {\color{black}transmit beamformer implemented via   one-bit phase shifters}. To solve the problem of  one-bit transmit beamforming, {we first convert the  tricky one-bit constraint into a continuous form}, and develop an efficient iterative method by applying the exact penalty method (EPM) \cite{le2015feature,yuan2016binary} and Nesterov-like gradient method \cite{6665045,7592408}. 

            \item Representative scenarios are considered to illustrate the performance of the proposed methods for the large-scale MIMO system  in terms of  the  relative entropy value and the detection performance in simulation. 
			
			%
			%
		\end{itemize}
	}

	
	\textit{Notation:} Vectors and matrices are denoted by lower case boldface letter $ \mathbf{a} $  and upper case  boldface letter $ \mathbf{A} $, respectively.    $ (\cdot)^T $ and $ (\cdot)^H $ represent the  transpose and conjugate transpose operators,  respectively. The sets of $n$-dimensional complex-valued  (real-valued) vectors and $ N \times N $ complex-valued   (real-valued) matrices are denoted by $\mathbb{C}^{n}$ ($ \mathbb{R}^{n} $) and $\mathbb{C}^{N\times N}$  ($ \mathbb{R}^{N\times N}$), respectively. $ \Re \left\{  \cdot  \right\} $ and $ \Im \left\{  \cdot  \right\} $ are typically reserved for the
	real part and the imaginary part of a complex-valued number,
	respectively.  $ {\rm vec}(\bf A) $   and ${\rm Tr}({\bf A})$ denote  the vectorization   and trace of $ \bf A $, respectively.     $ \mathbf{I}_{N} $ denotes the $ {N\times N} $ identity matrix.    $ \otimes $  and $\odot$  denote  the  Kronecker product and Hadamard product, respectively. The $ n $-th entry of $  \mathbf{a} $ is written as $ a(n) $. $ | \cdot |$ indicates  an  absolute value, cardinality, and determinant for a scalar  value, a set, and a matrix 
	depending on context.  

	\section{System Model}
\begin{figure}[t]
		\centering
		\includegraphics[width=0.85\linewidth]{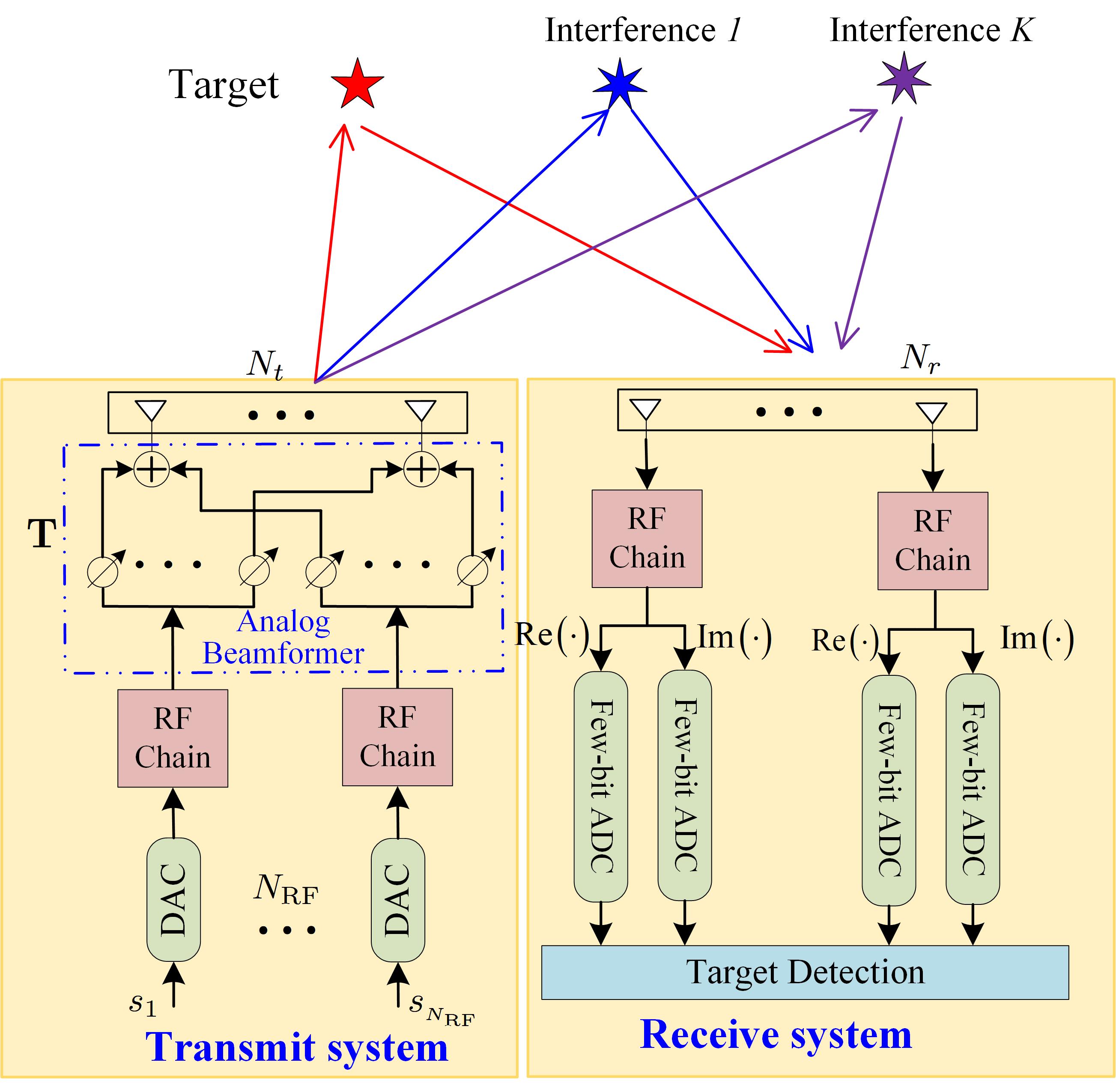}
		\vspace{-0.8em}
		\caption{{Overview of a large-scale MIMO system, where  the transmit and receive arrays are assumed to be colocated. The transmit array adopts a hybrid digital/analog architecture and the receive array adopts  few-bit ADCs.} }
		\vspace{-1em}
		\label{fig:pic1}
	\end{figure}

	We consider a  MIMO system  in which the transmitter is equipped with   $ N_t $ antennas and $ N_{\rm RF} $ RF chains.   
    The RF chains are connected to the antennas through an analog beamformer ${\bf T}\in {\mathbb{C}^{N_t\times N_{\rm RF}}}$, which is realized by
	analog phase shifters, we have $ |{\bf T}(i,j)|=\frac{1}{\sqrt{N_t}}   $.    
	The transmit and receive arrays are colocated, and the distance between the arrays
    to target/clutter is far larger than the spacing of the arrays
	such that a target located in the far field can be viewed as the same spatial angle relative to both of them\cite{4655353,5728938},  and arranged as half-wavelength spaced uniform linear arrays (ULAs).  
	While the receive system with $ N_r $ antennas  adopts a fully-digital structure, which means that each receive antenna is followed by an RF chain and a pair of  ADCs  for the real and imaginary parts of the received signal, as shown in Fig. \ref{fig:pic1}. For reducing the hardware cost, the ADCs are considered to be few-bit.  

	Denote the waveform of the $n$-th RF chain at the time instant $ \ell $ by $ s_{n}(\ell) $. Then the  transmitted waveform by the transmit array is given by 
	\[
	{\bf x}(\ell)=\sum\limits_{i = 1}^{{N_{{\rm{RF}}}}} {{{\bf{t}}_i}s_i(\ell)}=   {\bf T}{\bf s}(\ell) \in {\mathbb{C}^{N_t\times 1}}  , \ell=1, \cdots, L
	\]
	where  $ {\bf s}(\ell) =[s_1(\ell), \cdots, s_{N_{\rm RF}}(\ell)]^T $, {${\bf t}_i$  is the $i$-th column of the matrix $\bf T$, and it}  can be viewed as the  transmit beamspace weight corresponding to the $i$-th waveform \cite{5728938,6709762}.  In order to obtain the waveform diversity for the MIMO radar system, the orthogonality constraint $  {\bf T}^H{\bf T}={\bf I}_{N_{\rm RF}} $ has to be
	imposed \cite{5728938}. Besides, $L$ is the number of samples in the duration of the transmitted
	waveform. Moreover, we assume  that  {$ L \ge N_{\rm RF}$, and that  $   {\bf s}(\ell)    $ is normalized orthogonal waveform\footnote{In this work, the assumption $  L \ge N_{\rm RF} $ is necessary  for	applying the result ${\bf S}{\bf S}^H/L={\bf I}_{N_{\rm RF}}$, which implicitly indicate the condition for the proposed radar system to work.}, i.e., 
	${\bf S}{\bf S}^H/L={\bf I}_{N_{\rm RF}}$, where ${\bf S}=[{\bf s}(1), \cdots, {\bf s}(L)]$.}
	
	{\color{black}Suppose that there exists $ K $ stationary   clutters in the detection area},  then we can obtain   the received signals at the receive antennas  as 
	\begin{equation}
	\begin{aligned}
        {\bf y}(\ell)  = & { e^{\jmath 2 \pi (\ell - 1) f_{D,t} t_s }} {\bf H}{\bf T}{\bf s}(\ell) \\
        & +\sum\limits_{k = 1}^K  { e^{\jmath 2 \pi (\ell - 1) f_{D,k} t_s }} {\bf C}_{k} {\bf T} {\bf s}(\ell) + {\bf n}(\ell) 
	\end{aligned}
	\end{equation}
	where $ {\bf H}  $ is the target scattering matrix, for a collocated MIMO radar, the $\bf H$ becomes a rank-one matrix, as ${\bf H }= \xi {\bf a}_r(\theta_t){\bf a}_t^T(\theta_t)$, where $\xi$ is a random reflection coefficient,  and follows ${\cal CN}(0, \sigma_t^2 ) $, ${\bf a}_r(\theta)= \frac{1}{\sqrt{N_r}} {\left[ {{\rm{1, }}\; \cdots ,\;{e^{ - \jmath \pi ({N_r} - 1)\sin \theta }}} \right]^T}$ and ${\bf a}_t(\theta)= \frac{1}{\sqrt{N_t}} \left[ {{\rm{1, }}\; \cdots ,\;{e^{ - \jmath\pi ({N_t} - 1)\sin \theta }}} \right]^T$ respectively denote the
	receive and transmit steering vectors {\color{black}with $\theta_t$ being the direction of the target.}  
	$ {\bf{C}}_{k} =\xi_{c,k} {\bf a}_r(\theta_{c,k}){\bf a}_t^T(\theta_{c,k}) $  is the clutter response matrix of the $k$-th clutter, where $ \xi_{c,k} $ and $\theta_{c,k}$  separately denote  the reflection coefficient  and direction of the $k$-th clutter, {\color{black}we also assume that $   \theta_{c,1}, \cdots, \theta_{c,K}  $  are  fixed and generated according to a  uniform distribution of $[-\pi/2,\pi/2]$,  and that the number and directions of clutters are known in this model}. In addition, we also assume  that  $\left\lbrace  \xi_{c,k}\right\rbrace  $ are mutually independent and complex Gaussian
	random variables with zero mean and variance $\sigma_{c,k}^2  $. 
    $f_{D,t}$ and $f_{D,k}$ denote the Doppler frequencies of the target and $k$-th clutter, $t_s$ is the sample interval.
	$ {\bf n}(l) $ is a complex Gaussian random variables with zero mean and covariance matrix $\sigma_n^2 {\bf I}_{N_r}$.

	Concatenating $ L $ samples of received signals, we obtain 
	\begin{equation}
	{\bf Y}=   {\bf H}  {\bf T}{\bf S} { {\bf D}_0}  +  \sum\limits_{k = 1}^K  {\bf C}_{k}  {\bf T}{\bf S} { {\bf D}_{k}}   +  {\bf N}  \in {\mathbb C}^{N_r \times L}
	\end{equation}
	where ${\bf D}_0 = \mathrm{Diag}( [ 1 , \cdots ,  e^{\jmath 2 \pi (L - 1) f_{D,t}} ] ) $, 
    ${\bf D}_{k} = \mathrm{Diag}( [ 1 , \cdots ,  e^{\jmath 2 \pi (L - 1) f_{D,k}} ] ) $ , and 
    $ {\bf N}=[{\bf n}(1), \cdots, {\bf n}(L)] $.

	For the few-bit ADCs, we adopt an additive quantization noise model (AQNM) \cite{4407763,8640825}, which shows a reasonable accuracy  for Gaussian input signal \cite{4407763}. In the AQNM, the quantized output is linearized as a function of quantization bits $ B $. Thus, the quantized
	output signal at the receiver is expressed as 
	\begin{equation}
	\begin{aligned}
        {\bf Y}_q  &= {\cal Q}(  {\bf Y}   )  \\
        & \approx \alpha  {\bf H}  {\bf T}{\bf S} { {\bf D}_0} +  \alpha \sum\limits_{k = 1}^K  {\bf C}_{k}  {\bf T}{\bf S} { {\bf D}_k}  +  \alpha {\bf N} +{\bf N}_q  
	\end{aligned}
	\end{equation}
	where ${\cal Q}(\cdot)$ is the element-wise quantizer, $\alpha$ is  the quantization gain, defined as $\alpha =1-\beta$ with $ \beta $ being the normalized mean squared quantization error. For $ B=1, 2, 3, 4, 5 $, the corresponding values of $\beta$  are given by  0.3634, 0.1175, 0.03454, 0.009497, and  0.002499, respectively \cite{4407763}. {\color{black}$ {\bf N}_q $ is
	the additive Gaussian quantization noise introduced by the quantization operation, and is related to the input signal $ {\bf y}  $ and follows the  complex Gaussian distribution $ {\cal CN}({\bf 0}, {\bf R}_q ) $ with $ {\bf R}_q $ being given by \cite{4407763}}
	\begin{equation}
	\begin{aligned}
	{\bf R}_q  =\alpha \beta {\rm diag} \big( {\bf R}_{Y} \big)
	\end{aligned}
	\end{equation}
	where $  {\bf R}_{Y}  $ is the covariance matrix  of the $\bf Y$, given by  
	\begin{equation} 
	\begin{split}
        {\bf R}_{Y} = & L \sigma_t^2 \phi_t({\bf T}) {\bf a}_r (\theta_t) {\bf a}_r^H (\theta_t) \\ 
        & + L\sum\limits_{k = 1}^K \sigma_{c,k}^2 \phi_{c,k}({\bf T}){\bf a}_r (\theta_{c,k}) {\bf a}_r^H (\theta_{c,k})+ L\sigma _n^2{{\bf{I}}_{{N_r}}}
	\end{split}
	\end{equation}
	where $ \phi_t({\bf T}) \buildrel \Delta \over = {\bf a}_t^T(\theta_t) {\bf T}{\bf T}^H {\bf a}_t^*(\theta_t)$ and $  \phi_{c,k}({\bf T}) \buildrel \Delta \over = {\bf a}_t^T(\theta_{c,k}) {\bf T}{\bf T}^H {\bf a}_t^*(\theta_{c,k})$ denote the {\color{black}transmit power} for the directions of target and the $k$-th {\color{black}clutter}, respectively.

	Following \cite{8101018}, we establish the problem of detecting a
	target in the presence of observables by the following
	binary hypothesis test: 
	\begin{equation}  \left\{ \begin{array}{l}
	{{\cal H}_0}:\;{{\bf{Y}}_{q}} =    \alpha \sum\limits_{k = 1}^K  {\bf C}_{k}  {\bf T}{\bf S} { {\bf D}_k }  +  \alpha {\bf N}    +{\bf N}_{q0}  \\
	{{\cal H}_1}:\;{{\bf{Y}}_{q}} =\alpha  {\bf H}  {\bf T}{\bf S} { {\bf D}_0 }  +  \alpha \sum\limits_{k = 1}^K  {\bf C}_{k}  {\bf T}{\bf S} { {\bf D}_k}  +  \alpha {\bf N} +{\bf N}_{q1} 
	\end{array} \right. \label{8}
	\end{equation}
	where  the covariance matrices of the quantization noises $ {{\bf{N}}_{q0}}  $ and $ {{\bf{N}}_{q1}} $ are respectively given by 
	\begin{equation}
	\begin{aligned}
	& {\bf R}_{q0} =\alpha \beta L{\rm diag} \Big(\sum\limits_{k = 1}^K   \sigma_{c,k}^2 \phi_{c,k}({\bf T}) {\bf a}_r (\theta_{c,k}) {\bf a}_r^H (\theta_{c,k})+   \sigma _n^2{{\bf{I}}_{{N_r}}} \Big)\\
	&\quad ~~=\alpha \beta L\left( \sum\limits_{k = 1}^K \frac{\sigma_{c,k}^2}{N_r} \phi_{c,k}({\bf T}) + \sigma_n^2\right) {{\bf{I}}_{{N_r}}} 
	\end{aligned}
	\end{equation}
	and 
	\begin{equation}
	\begin{aligned}
	& {\bf R}_{q1}  =\alpha \beta L {\rm diag}\Bigg(   \sigma_t^2 \phi_t({\bf T}) {\bf a}_r (\theta_t) {\bf a}_r^H (\theta_t)+ \sum\limits_{k = 1}^K   \sigma_{c,k}^2 \phi_{c,k}({\bf T}) \\
	&\qquad \qquad \qquad \cdot {\bf a}_r (\theta_{c,k}) {\bf a}_r^H (\theta_{c,k})+  \sigma _n^2{{\bf{I}}_{{N_r}}}\Bigg)\\
	&\quad~~=\alpha \beta L\Bigg(  \frac{\sigma_{t}^2}{N_r}   \phi_t({\bf T}) + \sum\limits_{k = 1}^K  \frac{\sigma_{c,k}^2}{N_r} \phi_{c,k}({\bf T})  +  \sigma _n^2\Bigg){{\bf{I}}_{{N_r}}}
	\end{aligned}\label{8_1}
	\end{equation}
	
	
	In a radar system, the detection performance can be measured by the relative entropy, which is an information-theoretic metric \cite{8101018}.  
	Following \cite{8101018}, we select the  relative entropy as our criterion in this paper\footnote{As stated in Stein's Lemma in \cite{cover1999elements} that  for any given probability of false alarm $ P_{\rm fa} $, the 
		relative entropy is {\color{black}exponentially related} with the probability
		of miss  $ P_{\rm miss} $, i.e., $  {\rm{D(P_0||P_1)  = }}\mathop {{\rm{lim}}}\limits_{N \to \infty } \left( { - \frac{1}{N}\ln \left( {{P_{{\rm{miss}}}}} \right)} \right) $, which means that the maximization of  $ \rm{D(P_0||P_1) } $ leads to an asymptotic minimization of $ {{P_{{\rm{miss}}}}} $, i.e., maximization of the probability of detection $ P_{\rm d} $.}. Specifically, {\color{black}For a target direction    $\theta_t$ to be tested},  the relative entropy between the
	probability density functions  (PDFs)  associated with the two hypothesis can be computed as 
	\begin{equation}
	\begin{aligned}
	& D(P_0||P_1) \\
	& \quad=  \int {{f_0}({{\bf{Y}}_q})\log } \left( {\frac{{{f_0}({{\bf{Y}}_q})}}{{{f_1}({{\bf{Y}}_q})}}} \right)d{{\bf{Y}}_q}\\
	& \quad ={\mathbb E}_{f_0}\left\lbrace   \log {f_0}({{\bf{Y}}_q})    \right\rbrace  - {\mathbb E}_{f_0}\left\lbrace   \log {f_1}({{\bf{Y}}_q})    \right\rbrace\\
	& \quad = -\log \left|   {\bf R}_{Y0}  \right| +\log \left|   {\bf R}_{Y1}  \right| +{\rm Tr}\Big(   {\bf R}_{Y1}^{-1}  {\bf R}_{Y0}  \Big)-N_r
	\end{aligned}
	\label{9_1}
	\end{equation}
	where $ f_0({\bf Y}) $ and $f_1({\bf Y})$   separately denote PDFs corresponding to  the test hypotheses ${\cal H}_0$ and ${\cal H}_1$, 
	$ {\bf R}_{Y0}  $ and $ {\bf R}_{Y1}  $ are the  covariance matrices of $ {{\bf{Y}}_{q }} $ under the hypothesis $ {\cal H}_0 $ and $ {\cal H}_1 $, respectively, as 
	\begin{equation}
	\setlength{\abovedisplayskip}{3pt}
	\begin{aligned}
	 {\bf R}_{Y0}& =\alpha^2L\Big( \sum\limits_{k = 1}^K   \sigma_{c,k}^2 \phi_{c,k}({\bf T}) {\bf a}_r (\theta_{c,k}) {\bf a}_r^H (\theta_{c,k}) +  \sigma _n^2{{\bf{I}}_{{N_r}}} \Big) \\
	 &\quad  + {\bf R}_{q0}\\
	&= \alpha^2 L \Big( {\bf A}_c   {\bf \Phi}_c({\bf T}) {\bf A}_c^H   +   \sigma _n^2{{\bf{I}}_{{N_r}}} \Big)+ {\bf R}_{q0}
	\end{aligned}\label{10_1a}\\
	\end{equation}
	\begin{equation}
		\setlength{\abovedisplayskip}{3pt}
	\begin{aligned} 
	{\bf R}_{Y1} & = \alpha^2L\Bigg( \sigma_t^2 \phi_t({\bf T}) {\bf a}_r (\theta_t) {\bf a}_r^H (\theta_t)+ \sum\limits_{k = 1}^K   \sigma_{c,k}^2 \phi_{c,k}({\bf T})\\
	&\quad  ~~ ~~~~~   \cdot {\bf a}_r (\theta_{c,k}) {\bf a}_r^H (\theta_{c,k})+  \sigma _n^2{{\bf{I}}_{{N_r}}} \Bigg)
	+ {\bf R}_{q1} 
	\\
	&= \alpha^2 L \Big( {\bf A}_{tc}  {\bf \Phi}_{tc}({\bf T}) {\bf A}_{tc}^H   +   \sigma _n^2{{\bf{I}}_{{N_r}}} \Big)
 + {\bf R}_{q1}
	\end{aligned}\label{10_1b}
	\end{equation}
	where  $  {\bf A}_c  \buildrel \Delta \over = [ {\bf a}_r (\theta_{c,1}), \cdots, {\bf a}_r (\theta_{c,K})  ]$,  $  {\bf A}_{tc}   \buildrel \Delta \over = [ {\bf a}_r (\theta_t), {\bf A}_c   ]$,  $  {\bf \Phi}_c({\bf T}) \buildrel \Delta \over = {\rm diag}\big( [ \sigma_{c,1}^2 \phi_{c,1}({\bf T}),  \cdots,  \sigma_{c,K}^2 \phi_{c,K}({\bf T})   ]\big)$ and  $  {\bf \Phi}_{tc}({\bf T}) \buildrel \Delta \over = {\rm diag}\big( [ \sigma_t^2 \phi_t({\bf T}), \sigma_{c,1}^2 \phi_{c,1}({\bf T}), \cdots,  \sigma_{c,K}^2 \phi_{c,K}({\bf T})   ]\big)$.
	
	Plugging  \eqref{10_1a} and \eqref{10_1b}  into \eqref{9_1} yields $ D(P_0||P_1) $  given in \eqref{20_1} on the top of the next page. 
	\newcounter{MYtempeqncnt}
	\begin{figure*}[!t]
		\normalsize	
		\setcounter{MYtempeqncnt}{\value{equation}}
		\setcounter{equation}{11}	
		{   
			\begin{equation}
			\begin{aligned}
			D(P_0||P_1)= & \underbrace { \log \left| \alpha^2 L \Big( {\bf A}_{tc} {\bf \Phi}_{tc}({\bf T}) {\bf A}_{tc}^H  +   \sigma _n^2{{\bf{I}}_{{N_r}}} \Big)+ {\bf R}_{q1}   \right|}_{\buildrel \Delta \over = {\cal M}_a({\bf{T}})} 
			-\underbrace {\log \left| \alpha^2 L \Big( {\bf A}_c {\bf \Phi}_c({\bf T}) {\bf A}_c^H  +   \sigma _n^2{{\bf{I}}_{{N_r}}} \Big)+ {\bf R}_{q0}   \right|}_{\buildrel \Delta \over = {\cal M}_b({\bf{T}})}-N_r\\
			& + \underbrace {{\rm Tr}\Bigg( \Bigg(   \alpha^2 L \Big( {\bf A}_{tc} {\bf \Phi}_{tc}({\bf T}) {\bf A}_{tc}^H  +   \sigma _n^2{{\bf{I}}_{{N_r}}} \Big)+ {\bf R}_{q1}  \Bigg)^{-1}  \Bigg(\alpha^2 L \Big( {\bf A}_c {\bf \Phi}_c({\bf T}) {\bf A}_c^H  +   \sigma _n^2{{\bf{I}}_{{N_r}}} \Big)+ {\bf R}_{q0}\Bigg) \Bigg)}_{\buildrel \Delta \over = {\cal M}_c({\bf{T}})}
			\end{aligned}
			\label{20_1}
			\end{equation}}
		\hrulefill
		\setcounter{equation}{\value{MYtempeqncnt}}	
		\vspace{-3pt}
	\end{figure*}
	
	\addtocounter{equation}{1}

It is worthy to mention that optimizing relative entropy in \eqref{9_1} requires the prior knowledge of $\theta_t$ obtained by the cognitive paradigm.
However, from practical point of view, the exact knowledge of the
angle  of the target is {not  available}, and thus, it is reasonable to assume that $\theta_t$ is uniformly distributed random variable  with $ \theta_t \sim {\cal U}(\bar{\theta}_t -\Delta_t/2, \bar{\theta}_t +\Delta_t/2) $, where the mean $\bar{\theta}_t$ and  the range of uncertainty  $ \Delta_t $ are assumed to be known. In such case, the averaged relative entropy can be expressed as 
\begin{equation}
	{\bar{D}({P_0}||{P_1})} = \frac{1}{{{\Delta _t}}}\sum\limits_{{\theta _t} \in {\Theta _t}} {D({P_0}||{P_1})} 
\end{equation}
where the individual term  $ {D({P_0}||{P_1})} $ is given in \eqref{20_1},  the set $ {\Theta _t}    $ can be obtained by taking the discretization of the set $(\bar{\theta}_t -\Delta_t/2, \bar{\theta}_t +\Delta_t/2) $, i.e., 
\begin{equation}{\Theta _t} = \left[ {{{\bar \theta }_t} - \frac{{{\Delta _t}}}{2},\;{{\bar \theta }_t} - \frac{{{\Delta _t}}}{2} + {\delta _t},\;{{\bar \theta }_t} - \frac{{{\Delta _t}}}{2} + 2{\delta _t}, \cdots ,{{\bar \theta }_t} + \frac{{{\Delta _t}}}{2}} \right] \label{13}
\end{equation}
with $ \delta_t $ being the discrete spacing.
	
In this paper, we seek to design the constant-envelope beamformer $\bf T$ to maximize   the relative entropy. Concretely, our optimization problem can be formulated as follows
\begin{subequations}\label{16}
	\begin{align}
		&   {\bf T}=\arg{\max}  ~ {\bar D}(P_0||P_1)  \\
		&  \quad  \qquad {\rm{s}}.{\rm{t}}.    \left| {{\bf{T}} }(i,j)\right|  = \frac{1}{\sqrt{N_t}}, ~\forall i, j\\
		&  ~~~~~ \quad \qquad   {\bf{T}}^H{\bf{T}}={\bf I}_{N_{\rm RF}}
	\end{align}
\end{subequations}
	
	Notice that   the   problem \eqref{16} is nonconvex and difficult to    tackle   due to the complicated objective function. Towards that end, in the following, we will propose a two-stage method to    seek a suboptimal solution with low-complexity and satisfactory performance.

	\vspace{-0.7em}
	\section{The Proposed Two-Stage Method for Solving Problem \eqref{16}}
	{In this section, we shall present the two-stage  optimization framework  to design the constant-envelope beamformer.   To be more concretely,  the proposed optimization framework includes the two steps: 1) we optimize  the {\color{black}transmit power} for the directions of target and clutters, 
		and  2) we seek to design  the constant-envelope beamformer based on the obtained {\color{black}transmit power}.}
	
	\vspace{-0.7em}
	\subsection{Optimization of  the {\color{black}transmit power}}

Before maximizing the relative entropy with respect  to the {\color{black}transmit power} for the directions of target and clutters, the following lemma is useful.

\begin{lemma}\label{lem:1}
{Under the  assumption that $ \{\theta_k\}_{k=1}^{K+1} $ are independent random following a uniform distribution in $ [-\pi/2, \pi/2] $, we have that ${\bf a}_{r}^H(\theta_k) {\bf a}_{r}(\theta_k) =1$ and when $N_r\to \infty$, }
\begin{equation}
	\begin{aligned}
	&{\mathbb E}\left(  {\bf a}_{r}^H(\theta_k) {\bf a}_{r}(\theta_l) \right) \to 0, ~k\ne l   ~\\
	&  {\mathbb V}{\rm ar}\left( {\bf a}_{r}^H(\theta_k) {\bf a}_{r}(\theta_l)  \right) \to 0, k\ne l.
	\end{aligned}
\end{equation} 
\end{lemma}
\begin{IEEEproof}
	See Appendix \ref{prf:lem1}. 
\end{IEEEproof}

\begin{figure}[!t]
		\centering
        \vspace{-1em}
		\includegraphics[width=0.8\linewidth]{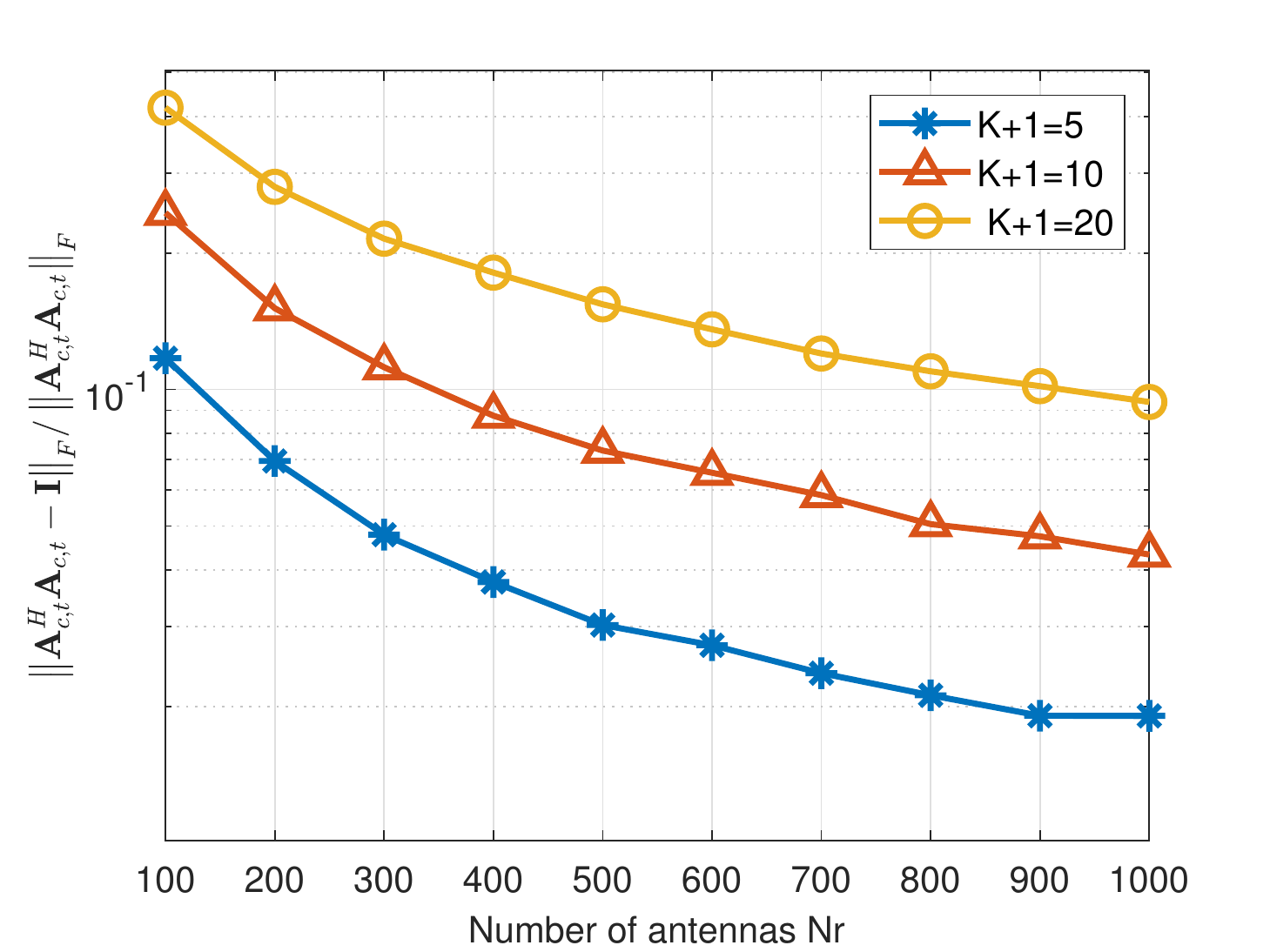}
        \vspace{-1em}
		\caption{Errors between the $ {\bf A}_{c,t}^H {\bf A}_{c,t} $ and $ {\bf I}_{K+1} $ versus the number of receive antennas $ N_r $ for different {numbers of clutters} $K$.}
		\label{fig:pic2_1}
        \vspace{-2em}
	\end{figure}

{
\begin{remark}
{\color{black}Lemma 1 implies that $ {\bf A}_{tc}^H {\bf A}_{tc} \to {\bf I}_{K+1} $, {as   $N_r \to \infty$}. Fig \ref{fig:pic2_1} shows the average difference between $ {\bf A}_{tc}^H {\bf A}_{tc} $ and $ {\bf I}_{K+1} $ over 10000 Monte Carlo simulations for each value of $ N_r $. We also find that the difference between  $ {\bf A}_{tc}^H {\bf A}_{tc} $ and $ {\bf I}_{K+1} $ becomes small as the number of clutters $K$ decreases.}
\end{remark}
\begin{remark}
{\color{black}
Strictly speaking, Lemma 1 is applied on the mean target angle $\bar{\theta}_t$ and $K$ clutter angles. Since the true target angle $\theta_t$ is uncertain within in the set $(\bar{\theta}_t -\Delta_t/2, \bar{\theta}_t +\Delta_t/2)$, we applied the result of Lemma 1 on the slightly shifted target angle (see \eqref{13}) and the other clutter angles. Considering that $\Delta_t$ is a small value closing to zero, $\sin(\bar{\theta}_t + \delta) \approx \sin(\bar{\theta}_t) + \epsilon$, and thereby the proof of Lemma 1 can still be valid with slight modification.
}
\end{remark}
}

Now, we consider the upper bound of {the first term of \eqref{20_1}}. Based on \eqref{8_1} and the identity $ \left|    {\bf I}+{\bf X}{\bf Y} \right| = \left|    {\bf I}+{\bf Y}{\bf X} \right| $, {the first term of \eqref{20_1}} can be converted to 
{\color{black}
\begin{equation}
	\begin{aligned}
	{\cal M}_a({\bf{T}}) 
	&={N_r}\log (\chi ) + \log \left| {{{\bf{I}}_{K + 1}} + \frac{{{\alpha ^2}L}}{\chi }{{\bf{\Phi }}_{tc}}({\bf{T}}){\bf{A}}_{tc}^H{{\bf{A}}_{tc}}} \right|\\
	&\approx{N_r}\log (\chi ) + \log \left| {{{\bf{I}}_{K + 1}} + \frac{{{\alpha ^2}L}}{\chi }{{\bf{\Phi }}_{tc}}({\bf{T}})} \right|\\
	&={N_r}\log (\chi ) + \log \left( {1 + \frac{{{\alpha ^2}L\sigma _t^2{\phi _t}}}{\chi }} \right)\\
	&~~~ + \sum\limits_{k = 1}^K \log\left( {1 + \frac{{{\alpha ^2}L\sigma _{c,k}^2{\phi _{c,k}}}}{\chi }}\right)
	\end{aligned}
\end{equation}}
where {\color{black}the second line is obtained by applying \textbf{Lemma 1} and the equality is achived when $N_r\rightarrow \infty$}, $\chi  = {\alpha ^2}L\sigma _n^2 + \alpha \beta L(\frac{{\sigma _t^2}}{{{N_r}}}{\phi _t} + \sum\limits_{k = 1}^K {\frac{{\sigma _{c,k}^2}}{{{N_r}}}} {\phi _{c,k}} + \sigma _n^2)$.
Here, for   notational simplification,  $ \phi_t({\bf T}) $ and  $ \phi_{c,k}({\bf T}) $ are separately abbreviated as  $  \phi_t $ and $  \phi_{c,k} $ in the remainder of this paper.
	
	Similarly, {the second term  of \eqref{20_1}} can be rewritten  as 
	{\color{black}\begin{equation}
	\begin{aligned}
	{\cal M}_b({\bf{T}}) &  
	\approx{N_r}\log (\varpi) + \log \left| {{{\bf{I}}_{K }} + \frac{{{\alpha ^2}L}}{\varpi}{{\bf{\Phi }}_{c}}({\bf{T}})} \right|\\
	&= {N_r}\log (\varpi) + \sum\limits_{k = 1}^K {\log \left( {1 + \frac{{{\alpha ^2}L\sigma _{c,k}^2{\phi _{c,k}}}}{\varpi }} \right)} 
	\end{aligned}
	\end{equation}}
	where $\varpi = {\alpha ^2}L\sigma _n^2 + \alpha \beta L( \sum\limits_{k = 1}^K {\frac{{\sigma _{c,k}^2}}{{{N_r}}}} {\phi _{c,k}} + \sigma _n^2)$.
	
	To proceed, we recast {the third term of \eqref{20_1}}, based on the matrix inversion lemma, we obtain that  
	\begin{equation} 
		\setlength{\abovedisplayskip}{3pt}
	\begin{aligned}
	&\Bigg(   \alpha^2 L \Big( {\bf A}_{tc} {\bf \Phi}_{tc}({\bf T}) {\bf A}_{tc}^H  +   \sigma _n^2{{\bf{I}}_{{N_r}}} \Big)+ {\bf R}_{q1}  \Bigg)^{-1}\\
	&\qquad={{\bf{D}}^{ - 1}} - {{\bf{D}}^{ - 1}}{{\bf{A}}_{tc}}{\left( {{\bf{I}} + {\alpha ^2}L{{\bf{\Phi }}_{tc}}{\bf{A}}_{tc}^H{{\bf{D}}^{ - 1}}{{\bf{A}}_{tc}}} \right)^{ - 1}}\\
&\qquad ~~~	\times \left( {{\alpha ^2}L{{\bf{\Phi }}_{tc}}} \right){\bf{A}}_{tc}^H{{\bf{D}}^{ - 1}}\\
	&\qquad=\frac{1}{\gamma }{{\bf{I}}_{{N_r}}} - \frac{1}{\gamma }{{\bf{A}}_{tc}} {\bf \Lambda}_{\bf T}  {\bf{A}}_{tc}^H
	\end{aligned}
	\end{equation}
	where $ {\bf D}=\gamma{\bf I}_{N_r}$ with  
	$
	\gamma= {\alpha ^2}L\sigma _n^2 + \alpha \beta L\Big(\frac{{\sigma _t^2}}{{{N_r}}}{\phi _t} + \sum\limits_{k = 1}^K {\frac{{\sigma _{c,k}^2}}{{{N_r}}}} {\phi _{c,k}} + \sigma _n^2\Big),
	$
	and 
	\begin{equation} 
	\begin{aligned}
	{\bf \Lambda_{\bf T}}=& {\rm diag}\Bigg( \Bigg[ \frac{{{\alpha ^2}L\sigma _t^2{\phi _t}}}{{\gamma {\rm{ + }}{\alpha ^2}L\sigma _t^2{\phi _t}}},\;\frac{{{\alpha ^2}L\sigma _{c,1}^2{\phi _{c,1}}}}{{\gamma {\rm{ + }}{\alpha ^2}L\sigma _{c,1}^2{\phi _{c,1}}}}, \\
	& \qquad \qquad \cdots, 	\frac{{{\alpha ^2}L\sigma _{c,K}^2{\phi _{c,K}}}}{{\gamma {\rm{ + }}{\alpha ^2}L\sigma _{c,K}^2{\phi _{c,K}}}} \Bigg] \Bigg)
	\end{aligned}
	\end{equation}
	
	Define $\eta  = {\alpha ^2}L\sigma _n^2 + \alpha \beta L( {\sum\limits_{k = 1}^K {\frac{{\sigma _{c,k}^2}}{{{N_r}}}} {\phi _{c,k}} + \sigma _n^2} )$,  the third part of $ D(P_0||P_1)(\theta_t) $ is rewritten in \eqref{21_1} on the top of the next page. 
	\begin{figure*}[!t]
		\normalsize	
		\setcounter{MYtempeqncnt}{\value{equation}}
		\setcounter{equation}{20}	   
\begin{equation} 
\begin{aligned}
{\cal M}_c({\bf{T}})    
& =   {\rm{Tr}}\left( \frac{1}{\gamma }\Big({{\bf{I}}_{{N_r}}} -  {{\bf{A}}_{tc}} {\bf \Lambda}_{\bf T}  {\bf{A}}_{tc}^H \Big) \Big( \eta {\bf I}_{N_r} +\alpha^2L{{\bf{A}}_c}{{\bf{\Phi }}_c}{\bf{A}}_c^H   \Big) \right)\\
& =  \frac{\eta }{\gamma } {\rm{Tr}}\left( {\bf{I}}_{N_r} - {{\bf{A}}_{tc}}  {\bf \Lambda_{\bf T}} {\bf{A}}_{tc}^H \right) 
+ \frac{{{\alpha ^2}L}}{\gamma }{\rm{Tr}}\left( {{{\bf{A}}_c}{{\bf{\Phi }}_c}{\bf{A}}_c^H} \right)  
 -\frac{{{\alpha ^2}L}}{\gamma }{\rm{Tr}}\left(  {{\bf{A}}_{tc}}  {\bf \Lambda_{\bf T}} {\bf{A}}_{tc}^H {{{\bf{A}}_c}{{\bf{\Phi }}_c}{\bf{A}}_c^H} \right)\\
& \approx \frac{\eta }{\gamma } N_r - \frac{\eta }{\gamma } {\rm{Tr}}\left(    {\bf \Lambda_{\bf T}}   \right)+ \frac{{{\alpha ^2}L}}{\gamma }{\rm{Tr}}\left(  {{\bf{\Phi }}_c} \right) 
 - \frac{{{\alpha ^2}L}}{\gamma }{\rm{Tr}}\left(  {\bf \Lambda_{\bf T}}  \left[ {\begin{array}{*{20}{c}}
	{\bf{0}}\\
	{{{\bf{I}}_K}}
	\end{array}} \right]{{\bf{\Phi }}_c} \left[ {\begin{array}{*{20}{c}}
	{\bf{0}}&{{{\bf{I}}_K}}
	\end{array}} \right] \right)\\
& = \frac{\eta }{{\gamma {\rm{ + }}{\alpha ^2}L\sigma _t^2{\phi _t}}} + \sum\limits_{k = 1}^K {\frac{{\eta  + {\alpha ^2}L  \sigma _{c,k}^2{\phi _{c,k}}  }}{{\gamma +  {\alpha ^2}L\sigma _{c,k}^2{\phi _{c,k}}}}}  + \frac{\eta }{\gamma }\left( {{N_r} - K - 1} \right)
\end{aligned}\label{21_1}
\end{equation}
		\hrulefill
		\setcounter{equation}{\value{MYtempeqncnt}}	
		\vspace*{-3pt}
	\end{figure*}
		\addtocounter{equation}{1}

	Now, we want to reformulate the problem \eqref{16} into a  {simpler}  one  with respect to $   \big\{   \Phi_t, \phi_{c,1}, \cdots, \phi_{c,K}   \big\} $ for large antenna arrays, where {\color{black}the  $\Phi_t$ denotes the transmit powers corresponding to possible directions of target $\Theta_t$ in \eqref{13}}. On the other hand, since ${\bf T}^H{\bf T}   =  {\bf I}_{\rm RF}$, one gets 
	\[
	{\bf a}_t^T(\theta_p) {\bf T}{\bf T}^H {\bf a}_t^*(\theta_p) \le \lambda_{\max} ({\bf T}{\bf T}^H)   =1, \theta_p \in \Theta_P
	\]
	where $\Theta_P=\{  \Theta_t, \theta_{c,1}, \cdots, \theta_{c,K}   \}$, {and $\lambda_{\max}(\cdot)$ denotes the maximum eigenvalue of a matrix}.

	With the above derivations, the  problem \eqref{16} with respect to $ { \Phi}_P  \buildrel \Delta \over =  \big\{   \Phi_t, \phi_{c,1}, \cdots, \phi_{c,K}   \big\} $ can be reformulated as
	\begin{subequations}\label{20}
		\begin{align}
		&     {  \{\phi_{p}\}  } =\arg{\max}  ~ g(  \{\phi_{p}\})  \buildrel \Delta \over = \sum\limits_{{\phi _t} \in {\Phi _t}} {\cal G}(\phi_t,  \{\phi_{c,k}\}) \\
		&  \quad  \qquad {\rm{s}}.{\rm{t}}. ~  0\le \phi_p\le 1, \forall  \phi_p   \in { \Phi}_P
		\end{align}
	\end{subequations}
	where the objective ${\cal G}(\phi_t,  \{\phi_{c,k}\})$ is defined in \eqref{21} on the top of the {next page}.
	\begin{figure*}[!t]
		\normalsize	
		\setcounter{MYtempeqncnt}{\value{equation}}
		\setcounter{equation}{22}	
		{   
			\begin{equation}
			\begin{aligned}
			{\cal G}(\phi_t, \{\phi_{c,k}\})  
			=&\log \left| {\left( {\alpha {\rm{ + }}\beta } \right)\sigma _n^2 + \left( {\alpha+ \frac{\beta }{N_r} } \right)\sigma _t^2{\phi _t}{\rm{ + }} \frac{\beta }{N_r} \sum\limits_{k = 1}^K {\sigma _{c,k}^2} {\phi _{c,k}}} \right| \\
			& +\sum\limits_{i =1 }^K \log \Bigg|   1+\frac{ {\beta } \sigma _t^2{\phi _t}/{N_r}}{ \left( {\alpha +\beta } \right)\sigma _n^2 +   \left( {\alpha +\frac{\beta }{N_r} } \right)\sigma _{c,i}^2{\phi _{c,i}}  
				+ \frac{\beta }{N_r}\sum\limits_{k \ne i}^K {\sigma _{c,k}^2{\phi _{c,k}}} }   \Bigg|\\
			&+\frac{\eta }{{\gamma {\rm{ + }}{\alpha ^2}L\sigma _t^2{\phi _t}}} + \sum\limits_{k = 1}^K {\frac{{\eta  + {\alpha ^2}L  \sigma _{c,k}^2{\phi _{c,k}}  }}{{\gamma {\rm{ + }}{\alpha ^2}L\sigma _{c,k}^2{\phi _{c,k}}}}}  + \frac{\eta }{\gamma }\left( {{N_r} - K - 1} \right)
			\end{aligned}
			\label{21}
			\end{equation}}
		\hrulefill
		\setcounter{equation}{\value{MYtempeqncnt}}	
		\vspace*{-3pt}
	\end{figure*}
	
	\addtocounter{equation}{1}
	Notice that the  complicated objective  function in problem \eqref{20} makes the problem  difficult to solve.  Fortunately, since the constraints in problem \eqref{20} are separable, the  block coordinate descent  (BCD)-type method can be exploited to   reach a stationary point of the problem \eqref{20} \cite{luo1992convergence}.  
	
	{At each iteration of the BCD method, a single
	variable is optimized, while the remaining variables
	are fixed.  More exactly, at the $t+1$-th iteration of the BCD method,  the optimization problem with respect to $ \phi_p, \forall p $ is written as} 
	\begin{equation}
	\begin{aligned}
	& \max\limits_{\phi_p}  ~ g (  \phi_{p},  {\bm \phi}_{-p}^{(t)}) ~~
	{\rm{s}}.{\rm{t}}. ~  0\le \phi_p\le 1, \forall  \phi_p   \in { \Phi}_P
	\end{aligned}\label{22}
	\end{equation} 
	where {\color{black} we define $   {\bm \phi}_{-p}^{(t)} =({  \phi}_{1}^{(t)},\cdots, {  \phi}_{p-1}^{(t)}, { \phi}_{p+1}^{(t)},\cdots, {  \phi}_{P}^{(t)} ) $}. We note that the objective function in  \eqref{22} is nonconvex and hard to deal with. {Actually, its suboptimal solution can be attained by using a one-dimensional search over $ (0,1) $.   Concretely, {we discretize the contiguous  range $ (0,1) $ as }
$ 	{\cal K}_p =[0, \delta_p, 2\delta_p,\cdots, 1] $, 
	where $\delta_p$ is the search interval. Thus, the   suboptimal $ {\phi_p} $ can be written as 
	\begin{equation}
	\hat{\phi}_p  = \arg \max \limits_{\phi_p \in {\cal K}_p} g \Big(  \phi_p ,  {\bm \phi}_{-p}^{(t)}\Big).
	\end{equation}}
	
	Note that   the main computational complexity of the BCD method  is linear with the number of iterations. In each iteration, we need to update $|\Phi_p|$ variables  successively.  Thus, the total number of multiplications for solving \eqref{20} is $ {\cal O}(I_1|\Phi_p|^2/\delta_p) $, where $ I_1 $ is {\color{black}the number of   iterations} of  the BCD. In addition, 
	since in each update of the proposed BCD, the objective function in \eqref{20}  is non-decreasing,  {\color{black}this algorithm is able to converge to  a stationary point}. 
	
	In the following, we need to design the constant-envelope beamformer based on the obtained $\{ \hat{\phi}_p\} $.
	\vspace{-1em}
	\subsection{Constant-Envelope Beamforming Design}
	In this subsection, we attempt to design the constant-envelope beamformer to make sure that   $  {\bf a}_t^T(\theta_p) {\bf T}{\bf T}^H {\bf a}_t^*(\theta_p), ~\theta_p \in \Theta_P $ {approaches}   the obtained $ \hat{\phi} (\theta_p) $. {\color{black}Towards that end}, the squared-error between
	the designed beampattern and the given beampattern is selected
	as the figure of merit, which is expressed as
	\begin{equation}
	\begin{aligned}
	{\rm MSE}({\bf T}) = 
	&  \sum\limits_{\theta_p \in \Theta_P } {|{\bf a}_t^T(\theta_p) {\bf T}{\bf T}^H {\bf a}_t^*(\theta_p)-\hat{\phi}(\theta_p) |^2} 
	\end{aligned}\label{22_1}
	\end{equation}
	Thus, our problem of the constant-envelope beamforming design with the orthogonality constraint can be formulated as 
	\begin{equation}
	\begin{aligned}
	& \min \limits_{\bf T}  ~ \sum\limits_{\theta_p \in \Theta_P} {|{\bf a}_t^T(\theta_p) {\bf T}{\bf T}^H {\bf a}_t^*(\theta_p)-\hat{\phi}(\theta_p)  |^2}  \\
	&  {\rm{s}}.{\rm{t}}. ~~  {\bf T}^H{\bf T}={\bf I}_{N_{\rm RF}}\\
	&\quad ~~~ \left| {{\bf{T}} }(i,j)\right|  = \frac{1}{\sqrt{N_t}}, \forall i, j
	\end{aligned} \label{25}
	\end{equation}
	Problem \eqref{25} is obviously non-convex due to
	the  orthogonality constraint. {To enforce a solution which meets the orthogonality requirement,  and simplify the
	problem}, we merge  the  orthogonality   constraint with the objective function, thus  problem \eqref{25}  {\color{black}can be relaxed} as 
	\begin{equation}
	\begin{aligned}
	& \min \limits_{\bf T}  {\cal Z}({\bf T})= \sum\limits_{\theta_p \in \Theta_P} {|{\bf a}_t^T(\theta_p) {\bf T}{\bf T}^H {\bf a}_t^*(\theta_p)-\hat{\phi}(\theta_p)  |^2}\\
	&\qquad \qquad \qquad   + \varsigma   \left\|  {\bf T}^H{\bf T}-{\bf I}_{N_{\rm RF}} \right\|^2_F   \\
	&  {\rm{s}}.{\rm{t}}. ~ \left| {{\bf{T}} }(i,j)\right|  = \frac{1}{\sqrt{N_t}}, \forall i, j
	\end{aligned} \label{25_1}
	\end{equation}
	where $\varsigma$ is a penalty parameter for the orthogonality   constraint\footnote{As for the penalty parameter $\varsigma$,    we select a very small $\varsigma$ to get a good performance point in the beginning,  then iteratively increase the $\varsigma$ the make sure that the  orthogonality condition is gradually enforced.}. 
	Nevertheless, problem
	\eqref{25_1} is still hard to solve directly due to the fact that  {the objective  function in problem \eqref{25_1} is nonconvex since variables are coupled inside the Frobenius norm and squared modulus}. 
	Fortunately, the MM 
	framework \cite{7547360,6601713} can be exploited   to reach a stationary point of problem \eqref{25_1}. The strategy for applying MM method is  to construct an accurate majorization function, instead of minimizing the function in problem \eqref{25_1}, the   majorization function is minimized at the $\color{black}(m)$-th iteration. 
	
	Before
	proceeding, the following useful Lemma is  given. 

    \begin{lemma}\label{lem:2}
	For a $ g( {\bf X}) =  [  {\bf u}^T  {\bf X}{\bf X}^H {\bf u}^*  ]^2$, we have that 
	\begin{equation}
	\begin{aligned}
&	g(  {\bf X}) \le  ( {{\bf{u}}^T}  { {\bf X}^{(m)} {\bf X}^{{(m)}H} }   {{\bf{u}}^*}  )    \left( {{\bf{u}}^T}   {\bf X}  {\bf X}^H  {\bf{u}}^* \right)   \\
	& -  \lambda _{\rm max}\left( {{\bf{u}}{{\bf{u}}^H} \otimes {{\bf{u}}^*}{{\bf{u}}^T}} \right)    \Re\big({\rm Tr} ( {\bf X}{\bf X}^H {\bf X}^{(m)} {\bf X}^{{(m)}H}) \big)  \\
	&+{\color{black}N_{\rm RF}{{{\color{black}\lambda _{\rm max}}\left( {{\bf{u}}{{\bf{u}}^H} \otimes {{\bf{u}}^*}{{\bf{u}}^T}} \right)}}}
	\end{aligned}
	\end{equation}
	where {$ {\bf X}^{(m)} $ is the obtained $\bf X$ at the ${(m)}$-th iteration},  {$ \lambda _{\rm max}({\bf A}) $ denotes the maximum   eigenvalue  of $\bf A$.} 
    \end{lemma}
    \vspace{-0.5em}
	\begin{IEEEproof}
		See  Appendix \ref{prf:lem2}.
	\end{IEEEproof}

	Based on Lemma 2 and the fact that $ \lambda _{\rm max}\left( {\bf{a}}_{t}({ \theta_p}){\bf{a}}_{t}^H({  \theta_p}) \otimes {\bf{a}}_{t}^*({ \theta_p}){\bf{a}}_{t}^T({ \theta_p}) \right) =1$, the following {\color{black}inequality} holds true 
	\begin{equation}
	\begin{aligned}
	{\cal Z}({\bf T}) \le {\rm Tr} \big( {\bf{T}} {\bf{T}}^H  {\bf Q}^{(m)}   \big) + {\color{black}\sum\limits_{\theta_p \in \Theta_P} {| \hat{\phi}(\theta_p)  |^2} + 2\varsigma N_{\rm RF} }
	\end{aligned}
	\end{equation}
	where the matrix $   {\bf Q}^{(m)}  $ is defined as 
	\begin{equation}
	\begin{aligned}
	&{\bf Q}^{(m)} \\
	&~~~=\sum\limits_{\theta_p \in \Theta_P} \left( {  {\bf{a}}_{t}^T({  \theta_p})  {\bf{T}}^{(m)}{\bf{T}}^{{(m)}H}{\bf{a}}_{t}^*({  \theta_p}) - 2 \hat{\phi}(\theta_p)  } \right){\bf{a}}_{t}^*({  \theta_p}) \\
	&~~~ \cdot {\bf{a}}_{t}^T({  \theta_p})-|\Theta_P|{\bf{T}}^{(m)}{\bf{T}}^{{(m)}H} + \varsigma \left(   {\bf{T}}^{(m)}{\bf{T}}^{{(m)}H}- 2{\bf I}_{N_t}     \right) 
	\end{aligned}\label{27}
	\end{equation}
${\bf T}^{(m)} $ is the obtained $\bf T$ at the ${(m)}$-th iteration. 
	
	With the above derivations, the majorization problem of   the constant-envelope beamforming design  can be written as 
	\begin{equation}
	\begin{aligned}
	& \min \limits_{\bf T}  ~ h({\bf T})= {\rm Tr} \big( {\bf{T}} {\bf{T}}^H  {\bf Q}^{(m)}   \big)  \\
	&  {\rm{s}}.{\rm{t}}. ~  \left| {{\bf{T}} }(i,j)\right|  = \frac{1}{\sqrt{N_t}}, \forall i, j
	\end{aligned}\label{23}
	\end{equation} 
	We note the objective function  $ h({\bf T})  $ is a quadratic form, to which
	a proper majorized function can be applied again.  According to the similar derivations in Appendix \ref{prf:lem2}, $ h({\bf T})  $ can be further majorized by the following
	function 
	\begin{equation}
	\begin{aligned}
	&{\rm Tr} \big( {\bf{T}} {\bf{T}}^H  {\bf Q}^{(m)}   \big)\\
	&\quad \le  h({\bf{T}}^{(m)} )+ \Re\big({ {\bf t}^{{(m)}H}}  ({\bf I}_{N_{\rm RF}}\otimes {\bf Q}^{(m)} )  ( {\bf t}-{\bf t}^{(m)} )\big)\\
	& \quad ~~ +\frac{{{\color{black}\lambda _{\rm max}} ({\bf I}_{N_{\rm RF}}\otimes {\bf Q}^{(m)} )}}{{\rm{2}}} \left\|  {\bf t} - {\bf t}^{(m)}\right\|^2 \\
	&\quad  =\Re\Big({ {\bf t}^{{(m)}H}}  \big({\bf I}_{N_{\rm RF}}\otimes {\bf Q}^{(m)} - {\color{black}\lambda _{\rm max}} ({\bf I}_{N_{\rm RF}}\otimes {\bf Q}^{(m)}  ){\bf I} \big)    {\bf t} \Big) \\
	&\qquad +{\color{black} N_{\rm RF}\lambda _{\rm max}  ({\bf I}_{N_{\rm RF}}\otimes {\bf Q}^{(m)}  )}
	\end{aligned}
    \label{eq:33}
	\end{equation}
	where ${\bf t}={\rm vec} ({\bf T})$.  Thus, problem \eqref{23} can be further majorized  as 
	\begin{equation}
	\begin{aligned}
	& \min\limits_{\bf t}  ~\Re\Big({ {\bf t}^{{(m)}H}}  \big({\bf I}_{N_{\rm RF}}\otimes {\bf Q}^{(m)} - {\color{black}\lambda _{\rm max}} ({\bf I}_{N_{\rm RF}}\otimes {\bf Q}^{(m)}  ){\bf I} \big)    {\bf t} \Big)\\
	&  {\rm{s}}.{\rm{t}}. ~  \left| {{\bf{t}} }(i)\right|  = \frac{1}{\sqrt{N_t}}, \forall i 
	\end{aligned}\label{24}
	\end{equation} 
	
	Let $ {{\bf{t}} }(i )=\frac{1}{\sqrt{N_t}} e^{\jmath \varphi_{i}}  $, then the problem  \eqref{24} has the
	following closed-form solution, as 
	\begin{equation}
	\varphi_{i} = \angle   {  \Big[ -\Big({\bf I}_{N_{\rm RF}}\otimes {\bf Q}^{(m)} - {\color{black}\lambda _{\rm max}}({\bf I}_{N_{\rm RF}}\otimes {\bf Q}^{(m)}  ){\bf I} \Big){ {\bf t}^{(m)}}  \Big]_i} 
	\label{35}
	\end{equation} 
	
	Now we discuss the complexity of the MM method for {designing} the constant-envelope beamformer, which is linear with the iteration number. In each iteration, we need  to compute the matrix $   {\bf Q}^m $   with a complexity of  $ {\cal O}\left(  N_{\rm RF}N_t^2\right)  $, and take a SVD of $  {\bf Q}^m  $ with  a complexity of $  {\cal O}\left(  N_{\rm RF}^3\right)   $.   
	Therefore, the
	overall complexity of  the proposed  MM method is   $  {\cal O} \left(I_2(  N_{\rm RF}N_t^2 +  N_{\rm RF}^3 )  \right) $, where $I_2$ is 
    {the number  of iterations of} the MM method. 
	
	{
    \begin{lemma}\label{lem:3}
        Assume that the sequence of the objective values generated by Algorithm 1 is $\{\mathcal{Z}^{(m)}\}$, then the sequence is non-increasing and will converge to a local minimum.
    \end{lemma}
    \vspace{-0.5em}
    \begin{IEEEproof}
        See Appendix \ref{prf:lem3}.
    \end{IEEEproof}
    
    The convergence rate of the MM method relies on the tightness of the majorization function, if the majorization function is a bad  upper bound to the original one, the convergence
	of the MM algorithm might be much slower. To speed up its convergence, we employ
	the accelerated scheme  based on the squared iterative method \cite{2010Simple}, which will not violate the convergence of the algorithm.
		
	{Let ${\mathbb{M}}$ denote the nonlinear updating map according to \eqref{35} during the iterations of the MM method}, and express the update of $ {\bf T}^{(m+1)} $ as  $ {\bf T}^{(m+1)}={\mathbb{M}}\left( {\bf T}^{(m)} \right)   $. Following the result in \cite{2010Simple},  we outline the main steps of  the  accelerated MM algorithm  for solving the  constant-envelope beamforming problem with the orthogonality constraint in Algorithm 1.
	}
	\begin{algorithm}
		\caption{The accelerated  MM method for   the constant-envelope  beamforming design with the orthogonality   constraint}
		\label{alg4_1}
		\begin{algorithmic}[1]
			\STATE \textbf{Input:}  Initializes $ {\bf T}^{(0)} $, the convergence parameter $\epsilon_{\rm AMM}>0$, $\varsigma>0$,  $\varpi >1$ and $M$.
			\STATE \textbf{Output:} $  {\bf T}   $.
			\STATE Set $ m=1 $.
			\REPEAT
			\STATE   $ {\bf Y}_1 = {\mathbb M}({\bf T}^{{(m-1)}})- {\bf T}^{(m-1)}$.    
			\STATE   $ {\bf Y}_2 =  {\mathbb M}\big( {\mathbb M}({\bf T}^{(m-1)}) \big)-{\mathbb M}({\bf T}^{(m-1)})- {\bf Y}_1$.   
			\STATE    $\kappa =- \| {\bf Y}_1\|_F/  \| {\bf Y}_2\|_F$.
			\STATE    $  {\bf Z} ={\bf T}^{(m)}-2\kappa {\bf Y}_1+ \kappa^2  {\bf Y}_2  $
			\STATE     $  {\bf T}^{(m)} = e^{\jmath \angle {\bf Z} }/\sqrt{N_t} $. 
			\STATE    Update  $  \varsigma=\varsigma\times \varpi $ every $M$ iterations. 
			\STATE  $ m=m+1 $.
			\UNTIL  $   { \left\|    {\bf T}^{(m)}-  {\bf T}^{(m-1)}  \right\|_F^2}   \le {\epsilon_{\rm{AMM}}} $.
		\end{algorithmic}
	\end{algorithm}

	\vspace{-0.5em}
	\section{Constant-Envelope Beamforming Design With  One-Bit Phase Shifters}
	The part of previous section considers the infinite-resolution
	phase shifters are available at the transmitter. However, in order to  {\color{black}maximally save  hardware  cost}, 
	in this subsection, we just consider the transmitter adopts  one-bit phase
	shifters, i.e., $ {\bf T}(i,j) \in  \frac{1}{\sqrt{N_t}}  \left\{ -1, ~1 \right\}, \forall i, j $, which can {maximally reduce the circuit power consumption and hardware cost.}

	Based on the  previous analysis, we design one-bit  transmit beamformer with orthogonality constraint {being written} as 
	\begin{equation}
	\begin{aligned}
	& \min \limits_{\bf T}    \sum\limits_{\theta_p \in \Theta_P} {|{\bf a}_t^T(\theta_p) {\bf T}{\bf T}^H {\bf a}_t^*(\theta_p)-\hat{\phi}_p |^2} +\varsigma\left\|  {\bf T}^T{\bf T}-{\bf I}_{N_{\rm RF}} \right\|^2_F  \\
	&  {\rm{s}}.{\rm{t}}. ~    {\bf T}(i,j) \in  \frac{1}{\sqrt{N_t}}  \left\{ -1, ~1 \right\}, \forall i, j 
	\end{aligned}\label{26}
	\end{equation} 
	The problem \eqref{26} can be solved {\color{black}by an exhaustive search} with exponential complexity $ {\cal O}(2^{N_t N_{\rm RF}}) $, which {\color{black}is not suitable for} large-scale system.  Alternatively, one may first obtain the solution
	by dropping  one-bit constraint,
	and then projecting the resulting solution onto a one-bit set. However, this method will suffer from a large performance loss. In the following, we will apply a variational reformulation of  one-bit constraint, and then propose {\color{black}an efficient method inspired by the idea of the exact penalty method (EPM)}\cite{le2015feature,yuan2016binary} to solve the reformulated one-bit beamforming problem.

	To solve the problem  \eqref{26}, {\color{black}the following proposition is useful.}
	\begin{proposition}\label{pro:1}
	One-bit set $ \left\lbrace  {\bf x} \in \frac{1}{\sqrt{M}}  \left\{ -1, ~1 \right\}^{N} \right\rbrace   $ is equivalent to the set 
	\[ {\Psi}= \left\lbrace  ({\bf x}, {\bf y})| {\bf x}^T{\bf y}=\frac{N}{M}, -\frac{1}{\sqrt{M}} \le {\bf x}\le \frac{1}{\sqrt{M}},  {\bf y}^T{\bf y} \le \frac{N}{M} \right\rbrace. \]     
    \end{proposition}
	\begin{IEEEproof}
		See  Appendix \ref{prf:pro1}.
	\end{IEEEproof}

	Based on the Proposition 1, one-bit problem \eqref{26} can be reformulated as the following continuous one
	\begin{equation}
	\begin{aligned}
	& \min \limits_{\bf t,~v}  ~ \sum\limits_{\theta_p \in \Theta_P} {( {\bf t}^T {\bf \Phi}_p{\bf t}  -\hat{\phi}_p )^2} +\varsigma  \left\|  {\bf T}^T{\bf T}-{\bf I}_{N_{\rm RF}} \right\|^2_F\\
	&  {\rm{s}}.{\rm{t}}. ~   -\frac{1}{\sqrt{N_t}} \le {\bf t}\le \frac{1}{\sqrt{N_t}}\\
	&\quad ~~ ~  {\bf v}^T{\bf v} \le  N_{\rm RF},~  {\bf t}^T{\bf v}  = N_{\rm RF},  
	\end{aligned} \label{27_1}
	\end{equation} 
	where ${ {\bf \Phi}_p =\Re\Big( {\bf I}_{N_{\rm RF}} \otimes  {\bf{a}}_{t}^*({  \theta_p}) {\bf{a}}_{t}^T({  \theta_p})\Big) }$ and {{\bf t}={\rm vec}({\bf T})}. In the following, we will solve problem \eqref{27_1} by utilizing the EPM, which penalizes the complementary
	error directly by a penalty function. 
	
	More exactly,  {problem}  \eqref{27_1}  is expressed as
	\begin{equation}
	\begin{aligned}
	& \min \limits_{\bf t,~v}  ~ \sum\limits_{\theta_p \in \Theta_P} {( {\bf t}^T {\bf \Phi}_p{\bf t}  -\hat{\phi}_p )^2} +   \rho(N_{\rm RF}-{\bf t}^T{\bf v})\\
	&\quad \qquad     +\varsigma \left\|  {\bf T}^T{\bf T}-{\bf I}_{N_{\rm RF}} \right\|^2_F \\
	&  {\rm{s}}.{\rm{t}}. ~   -\frac{1}{\sqrt{N_t}} \le {\bf t}\le \frac{1}{\sqrt{N_t}},~{\bf v}^T{\bf v} \le  N_{\rm RF}  
	\end{aligned} \label{28}
	\end{equation} 
	where $\rho$ is the penalty parameter for the constraint $ {\bf t}^T{\bf v}  = N_{\rm RF} $.  Note that $N_{\rm RF}-{\bf t}^T{\bf v} \ge 0 $  always  holds for any feasible $  ({\bf t}, {\bf v}) $. The selection scheme of $\rho$ is similar to that of $\varsigma$.  
	{To solve the problem \eqref{28} with a fixed $\rho$,    we can first obtain the closed-form solution of $\bf v$, which is dependent on $\bf t$, and   substituted it into problem \eqref{28}, and update $\bf t$ with aid of the gradient method. 
	
 Specifically,  for an arbitrary   ${\bf t} $, the $\bf v$ is updated by solving 
 \begin{equation}
 \begin{aligned}
 & \min \limits_{\bf v }  ~  -{\bf t}^T {\bf v}, ~~   
    {\rm{s}}.{\rm{t}}. ~ ~~   {\bf v}^T{\bf v} \le  N_{\rm RF}
 \end{aligned} \label{30}
 \end{equation} 
 whose closed-form solution is  $ {\bf v} = {{\sqrt {{N_{{\rm{RF}}}}} {{\bf{t}} }}}/{{\left\| {{{\bf{t}} }} \right\|}} $.

	 Substituting $ {\bf v} = {{\sqrt {{N_{{\rm{RF}}}}} {{\bf{t}} }}}/{{\left\| {{{\bf{t}} }} \right\|}} $ into  problem \eqref{28}, problem \eqref{28} reduces to a minimization problem with respect to $\bf t$, as 
	\begin{equation}
	\begin{aligned}
	& \min \limits_{\bf t }  ~ {\cal F}_{\rho}({\bf t} )=\sum\limits_{\theta_p \in \Theta_P} {( {\bf t}^T {\bf \Phi}_p{\bf t}    -\hat{\phi}_p )^2}  +\rho(N_{\rm RF}- \sqrt{N_{\rm RF}} \|{\bf t} \|)\\
	& \quad \qquad \qquad \quad \quad    +\varsigma \left\|  {\bf T}^T{\bf T}-{\bf I}_{N_{\rm RF}} \right\|^2_F\\
	&  {\rm{s}}.{\rm{t}}. ~   -\frac{1}{\sqrt{N_t}} \le {\bf t}\le \frac{1}{\sqrt{N_t}} 
	\end{aligned} \label{29}
	\end{equation} 
	The problem \eqref{29} is a large-scale smooth problem with a compact constraint, to take advantage of {the structure} of the constraint, we develop an accelerated proximal gradient  method, i.e., Nesterov-like gradient  method \cite{6665045,7592408}, to handle it.  To be more specific, at the $\color{black}(p)$-th iteration, the Nesterov-like gradient  method updates $\bf t$ as follows
	\begin{subequations} 
		\begin{align}
		&    {\bm w}^{(p)}= {\bf t}^{(p)}+\frac{\tau^{(p-1)}}{\tau^{(p+1}}\Big( {\bf t}^{(p)}-{\bf t}^{(p-1)}  \Big)\\
		&    {\bf t}^{(p+1)}= \prod\nolimits_{ - \frac{1}{{\sqrt {{N_t}} }} \le {\bf{t}} \le \frac{1}{{\sqrt {{N_t}} }}} \Big({\bm w}^{(p)}-  \mu^{(p)} \nabla_{\bf t} {\cal F}_{\rho}({\bf t} )   \Big) 
		\end{align}
	\end{subequations}
	where $\tau^{(p+1)} =\frac{{1 + \sqrt {1 + 4{{\left( {{\tau ^{(p)}}} \right)}^2}} }}{2} $, $ \mu^{(p)} $ is a step size which can be
	determined by a backtracking line search \cite{boyd2004convex},   $ \prod\nolimits_{{\cal S}}(\cdot) $ is the projection operator onto set $ \cal S $. Additionally, as derived in Appendix \ref{prf:grad}, the $(j-1)\times N_t+i$th element of  $ \nabla_{\bf t} {\cal F}_{\rho}({\bf t} ) $ is given by 
	\begin{equation}
	\begin{aligned}
	&[\nabla_{\bf t} {\cal F}_{\rho}({\bf t} ) ]_{(j-1)\times N_t+i}=\\
	&  \sum\limits_{\theta_p \in \Theta_P} {({{\bf{t}}^T}{{\bf{\Phi }}_p}{\bf{t}} - {{\hat \phi }_p}){{\bf{\Phi }}_{p,((j - 1){N_t} + i):}}{\bf{t}}}  -   \frac{\rho \sqrt{N_{\rm RF}}}{\|{\bf t}\| }t_{(j - 1){N_t} + i}  \\
	&~~ +\varsigma \Bigg( 4T_{i,j}^3+4T_{i,j}\left( \big[{\overline {\bf{T}} }_{i,j}^T  {\overline {\bf{T}} }_{i,j}  -{\bf I}_{N_{\rm RF}}  \big]_{j,j}+ \big[  {\overline {\bf{T}} }_{i,j} {\overline {\bf{T}} }_{i,j}^T   \big]_{i,i}  \right)   \\
	&~~~~ ~~~~~~ +4 \big[ \overline {\bf{T}}  _{i,j} \big(  {\overline {\bf{T}} }_{i,j}^T  {\overline {\bf{T}} }_{i,j}  -{\bf I}_{N_{\rm RF}}    \big)   \big]_{i,j}  \Bigg),\\
	&\qquad \quad j=1,\cdots, N_{\rm RF}; i=1,\cdots, N_t
	\end{aligned} 
	\end{equation}
	where $  {{{\bf{\Phi }}_{p,((j - 1){N_t} + i):}}}  $ denotes the $(j - 1){N_t} + i  $-th row of the matrix $ {\bf{\Phi }}_p$,  $ t_{(j - 1){N_t} + i}  $ stands for the  $(j - 1){N_t} + i  $-th entry of the vector $ {\bf t}  $, and $  {\overline {\bf{T}} }_{i,j} $ is the matrix $\bf{T}  $  whose $ (i, j) $-th entry is
	zeroed.
	
	We note that  the Nesterov-like gradient  method provides a
	faster convergence {rate    than} that of the traditional
	gradient method \cite{6665045,7592408}.

	Based on the above analysis, the Nesterov-like gradient  method  for   one-bit  beamforming design is summarized in Algorithm 2.

		The main complexity of the proposed method for one-bit transmit beamforming design is  caused by computing the gradient of $ {\cal F}_{\rho}({\bf t} ) $, i.e., $ \nabla_{\bf t} {\cal F}_{\rho}({\bf t} ) $, which requires a  complexity of $  {\cal O} \left(  |\Phi_p|  N_t^2N_{\rm RF}^2    \right) $.   Therefore, the
		overall complexity of  one-bit beamforming design  is   $  {\cal O} \left(  {  I}_3 |\Phi_p| N_t^2N_{\rm RF}^2     \right) $, where $I_3$ is {the
		number of iterations}  of  the Nesterov-like gradient  method.
	}

	\begin{algorithm}[t]
		\caption{The Nesterov-like gradient  method  for  designing one-bit  beamformer.}
		\label{alg5_1}
		\begin{algorithmic}[1]{
			\STATE \textbf{Input:}  Initializes  $ {\bf t}^{(0)}, \tau^{(0)}$, the convergence parameter $\epsilon_{\rm EPM}>0$, $\varsigma, \rho>0$,  $\varpi,\varrho   
			>1$ and $M, N$.
			\STATE \textbf{Output:} $  {\bf t}   $.
			\STATE Set $p=0 $.
			\REPEAT
			\STATE   Compute $ {\tau}^{{(p+1)}}  = \tau^{{(p+1)}} =\frac{{1 + \sqrt {1 + 4{{\left( {{\tau ^{(p)}}} \right)}^2}} }}{2}$.     
			\STATE   Obtain $ \mu^{(p)} $  by a backtracking line search.
			 \STATE   Compute $ {\bm w}^{(p)}= {\bf t}^{(p)}+\frac{\tau^{(p)}-1}{\tau^{(p+1)}}\Big( {\bf t}^{(p)}-{\bf t}^{(p-1)}  \Big) $.    
			 	\STATE    Update  $    {\bf t}^{(p+1)}= \prod\nolimits_{ - \frac{1}{{\sqrt {{N_t}} }} \le {\bf{t}} \le \frac{1}{{\sqrt {{N_t}} }}} \Big({\bm w}^{(p)}-  \mu^{(p)} \nabla_{\bf t} {\cal F}_{\rho}({\bf t} )   \Big)  $.  
			\STATE    Update  $   \varsigma =\varsigma \times \varpi $ every $M$ iterations. 
			\STATE    Update  $  \rho =\rho \times \varrho  $ every $N$ iterations.
			\STATE    Compute  $ {\bf g}^{(p+1)} =\nabla_{\bf t} {\cal F}_{\rho}({\bf t}^{(p+1)} )    $.
			\STATE  $ p=p+1 $.
			\UNTIL  $   { \left\|    {\bf g}^{(p)}   \right\| }   \le {\epsilon_{\rm{EPM}}} $.}
		\end{algorithmic}
	\end{algorithm}

	\section{Numerical Simulations}
	In this section, several sets of numerical simulations are
	presented to assess the performance of the proposed  beamforming designs.  
	{\color{black}We first evaluate the performance of  the constant-envelope beamformer
	with infinite-resolution phase shifters}, and then the  beamforming design with  one-bit  phase shifters  is considered.

	Unless otherwise specified, in all simulations,  {we consider that the} transmitter and
	receiver are ULAs of  $ N_t=128 $ and $N_r=128$, respectively. 
	The transmit array   is equipped with  $N_{\rm RF}=8$ RF chains.  The transmit waveform  $ \bf S $ is chosen to be orthogonal LFM \cite{8141978}, whose code length is $L=16$.  We assume the target is located at $\theta_t=0^\circ$ with the uncertainty of ${\Delta_t =2^\circ}$, and the variance of its reflection coefficient  is $\sigma_t^2=0$ dB. In addition, we assume that the scattering
	strength is   $ \sigma^2_{c,k}=30 $ dB (for all $k$).
 The directions of the clutter  considered in our simulations are listed in the following table.

	\renewcommand\arraystretch{1.2}
\begin{table*}[t]
	\caption{The parameter settings of the clutters in simulations.}
	\label{table1}
	\setlength{\tabcolsep}{4pt}
	\begin{center}
		\begin{tabular}{|c |c    |}
			\hline 	
			& The directions of the clutters	\\
			\hline	
		   $K=5$		   & $-31.0^\circ,  -3.3^\circ, 28.7^\circ, -73.7^\circ, 69.1^\circ $\\
		   	\hline	
			$K=10$
		 	   & $-15.3^\circ,22.0^\circ, 59.9^\circ, -47.6^\circ,-75.4^\circ,78.9^\circ , -11.4^\circ,61.9^\circ,34.1^\circ, -64.7^\circ$ \\
			\hline
					$K=15$ 		   & $-40.4^\circ, -12.1^\circ, 24.6^\circ, -16.7^\circ, -72.0^\circ, 30.4^\circ, 47.5^\circ, 75.8^\circ, -86.2^\circ, -12.7^\circ, -79^\circ,$\\ & $-81.9^\circ, 54.4^\circ, 22.9^\circ,
		3.6^\circ $  \\
			\hline
		$K=20$ 		   & $34.1^\circ, -4.3^\circ, -9.0^\circ, -42.6^\circ, 39.5^\circ, 89.6^\circ, -12.4^\circ, -23.1^\circ, 25.2^\circ, 8.7^\circ, 45.1^\circ,$\\ & $-82.9^\circ, 34.1^\circ, -6.2^\circ,
		15.2^\circ, 2.4^\circ, 22.5^\circ, -18.7^\circ, -33.6^\circ, 32.3^\circ $  \\
			\hline	   
		\end{tabular}
	\end{center}
\end{table*}
	
	The variance of the Gaussian white noise is $  \sigma_n^2= 0
	$ dB.  As to the stop criteria of the proposed algorithms, we
	set   the tolerance $ \epsilon=10^{-4} $. 

	In all simulations, we denote the MM method with acceleration and the general MM without acceleration as ``AMM",   and ``MM", respectively. 
	
	\subsection{Constant-Envelope beamforming with infinite-resolution phase shifters}
 In this subsection, we  examine  the performance of the proposed beamforming method   with infinite-resolution phase shifters. 
 
\begin{figure}[!t]
		\centering
		\subfigure[]{
			\label{fig:subfig:1a} 
			\includegraphics[width=0.8\linewidth]{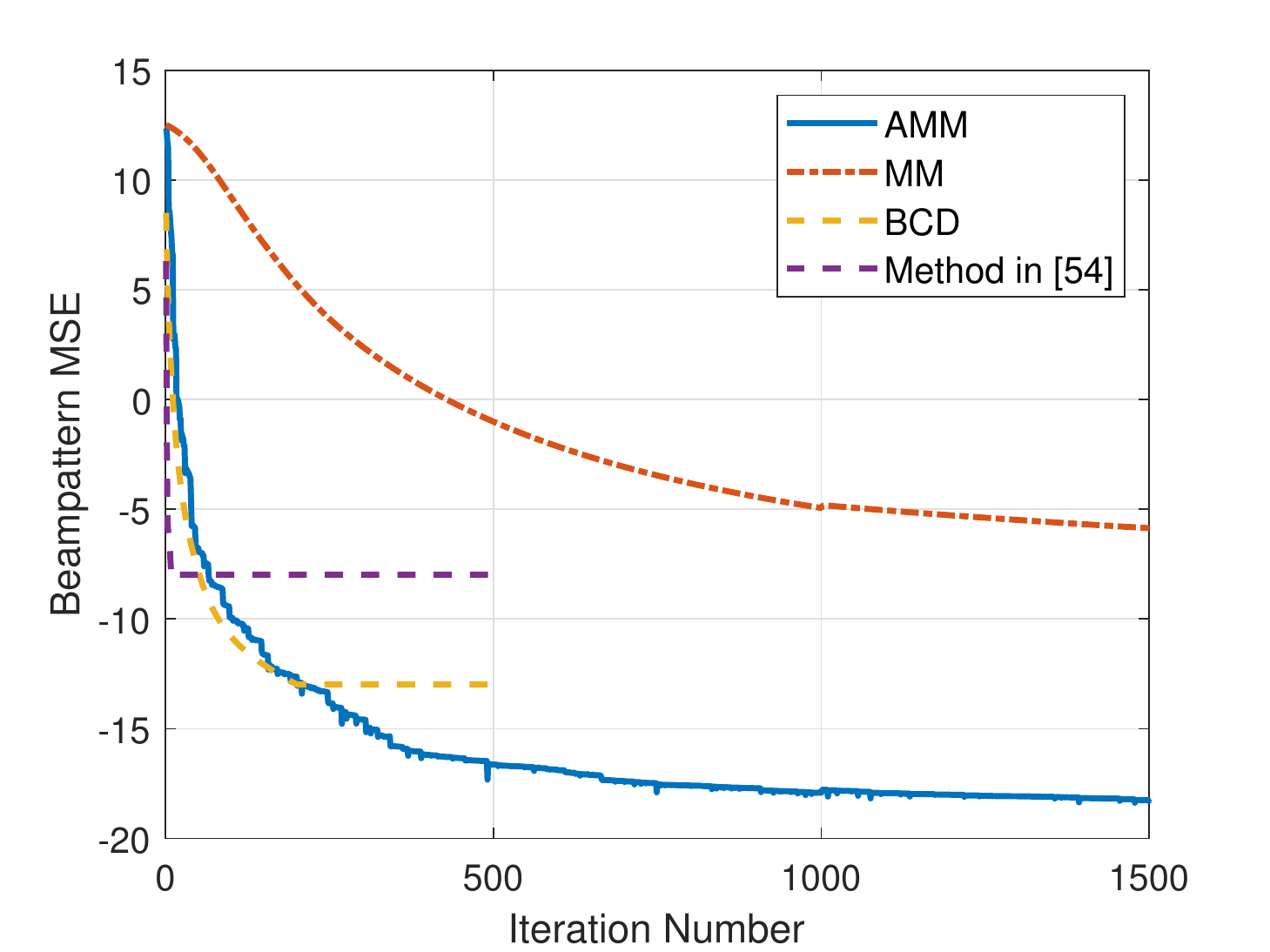}}
		\hspace{-0.2in}
		\subfigure[]{
			\label{fig:subfig:1b} 
			\includegraphics[width=0.8\linewidth]{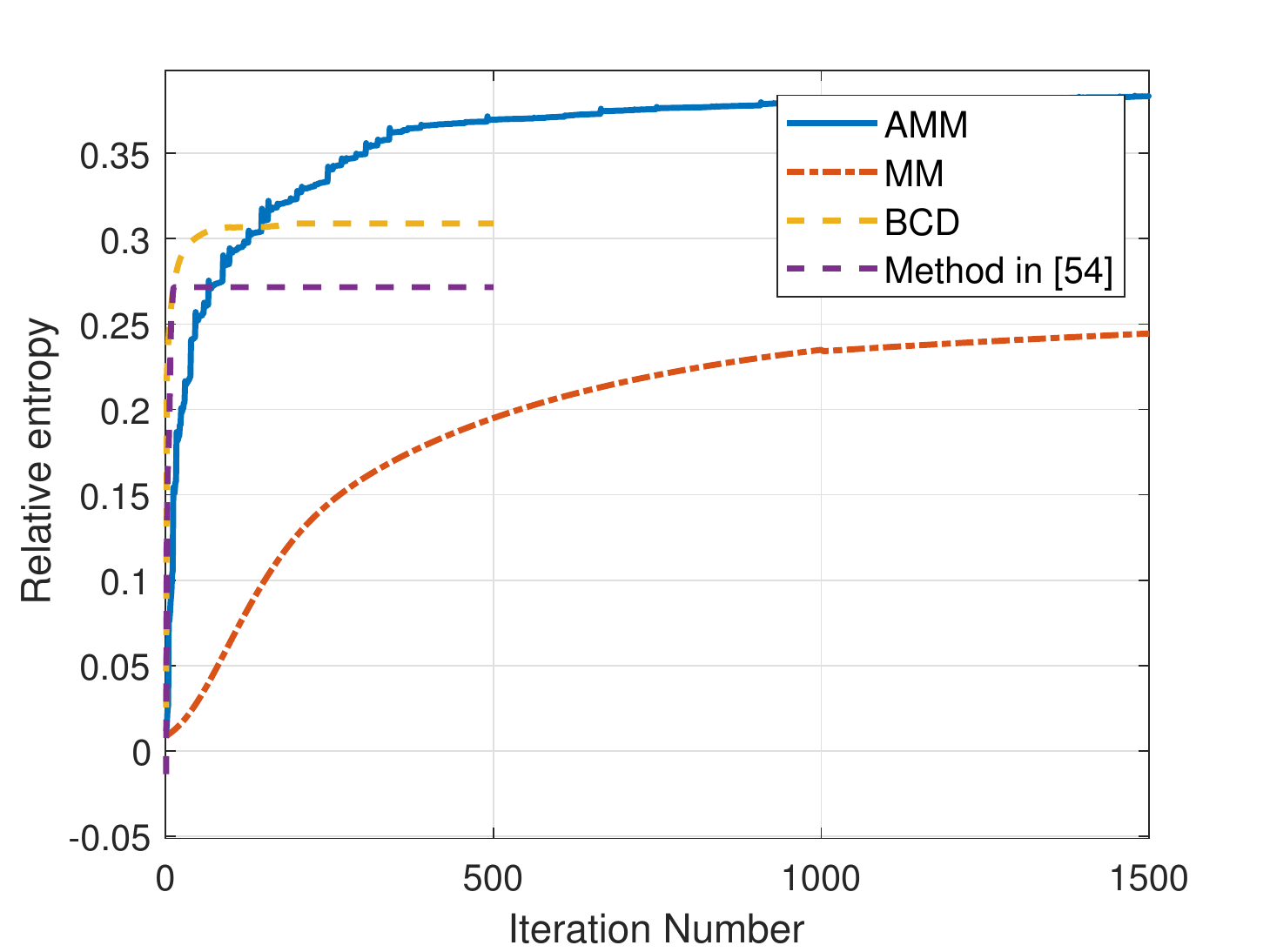}}
		\caption{The convergence performance of the proposed constant-envelope beamforming method. (a) The  beampattern MSE of the designed constant-envelope beamformers  versus iteration number, (b) the relative entropy of the designed constant-envelope beamformers  versus iteration number.		\vspace{-1em}}
		\label{fig:pic11} 
		\vspace{-1em}
	\end{figure}

	{\textit{Example 1:}}	Fig. \ref{fig:pic11} analyzes   
	the convergence performance of the MM method with acceleration for solving problem \eqref{22_1} by considering one-bit ADCs adopted at the receiver,  $N_t=128$,  $N_{\rm RF}=8
	$ and $N_r=128$. {\color{black}We consider a scenario where the $10$ clutter scatterers, their directions  are randomly generated  with a uniform  distribution over $ [-90^\circ, 90^\circ] $, to be specific,  the  considered directions of clutters are listed in Tab. I.} Fig. \ref{fig:pic11}(a) and Fig. \ref{fig:pic11}(b) respectively  reveal  the {beampattern MSE  (i.e., $ {\rm MSE}({\bf T})$ in Eq. \eqref{22_1})}  and relative entropy of the designed beamformer by using the AMM method, for comparison purpose, the general MM,  BCD and {the  two-stage  method in \cite{8359370} that firstly  obtains the optimal digital beamformer
	by dropping the nonconvex  constant-envelope constraint,
	and then projects the optimal digital beamformer  onto the constant modulus   set} are also considered. The result shows that  the convergence	rate of the general MM is	rather slower  
	than that of the AMM, {this result agrees with that in \cite{2010Simple}}. {Moreover, {the proposed AMM}  {method remarkably outperforms the MM, BCD  and {\color{black} the two-stage  method in \cite{8359370}}  in terms of the relative entropy performance}}.  The corresponding relative entropy,  computational complexity per iteration and the total  iterations number
	as well as the {\color{black}consumed}  time per iteration are given
	in Table II. The results in Table II also show that the AMM method  performs better than the other two methods 
	with respect to  the relative entropy and computational efficiency. 
	\begin{table*}[t]
		\centering
		\fontsize{8}{11}\selectfont
		\caption{Property analysis of various methods}
		\label{tab:2}
		\begin{tabular}{|c|c|c|c|c|}
			\hline
			\multirow{1}{*}{}&
			\multicolumn{1}{c|}{AMM}&\multicolumn{1}{c|}{MM}&\multicolumn{1}{c|}{BCD}&\multicolumn{1}{c|}{Method in [54]}\cr\cline{2-3}
			\hline
			\hline
			relative entropy& 0.3833 & 0.2439& 0.3071& 0.2713\cr\hline
			computational complexity&${\cal O}(  N_{\rm RF}N_t^2 +  N_{\rm RF}^3 )$& ${\cal O}(  N_{\rm RF}N_t^2 +  N_{\rm RF}^3 )$ & ${\cal O}(  N_{\rm RF}^2N_t^3 K +N_t^3N_{\rm RF}K^2)$ & ${\cal O}(  N_{\rm RF}N_t^2 K +N_{\rm RF}K^2)$\cr\hline
			total  iterations number&  1500 &1500 & 500 & 500\cr\hline
			{\color{black}consumed} time per iteration & 0.0563&0.0378 & 3.241 & 0.5062\cr\hline
		\end{tabular}
	\end{table*}
	
\begin{figure}[!t]
	\centering
	\includegraphics[width=0.8\linewidth]{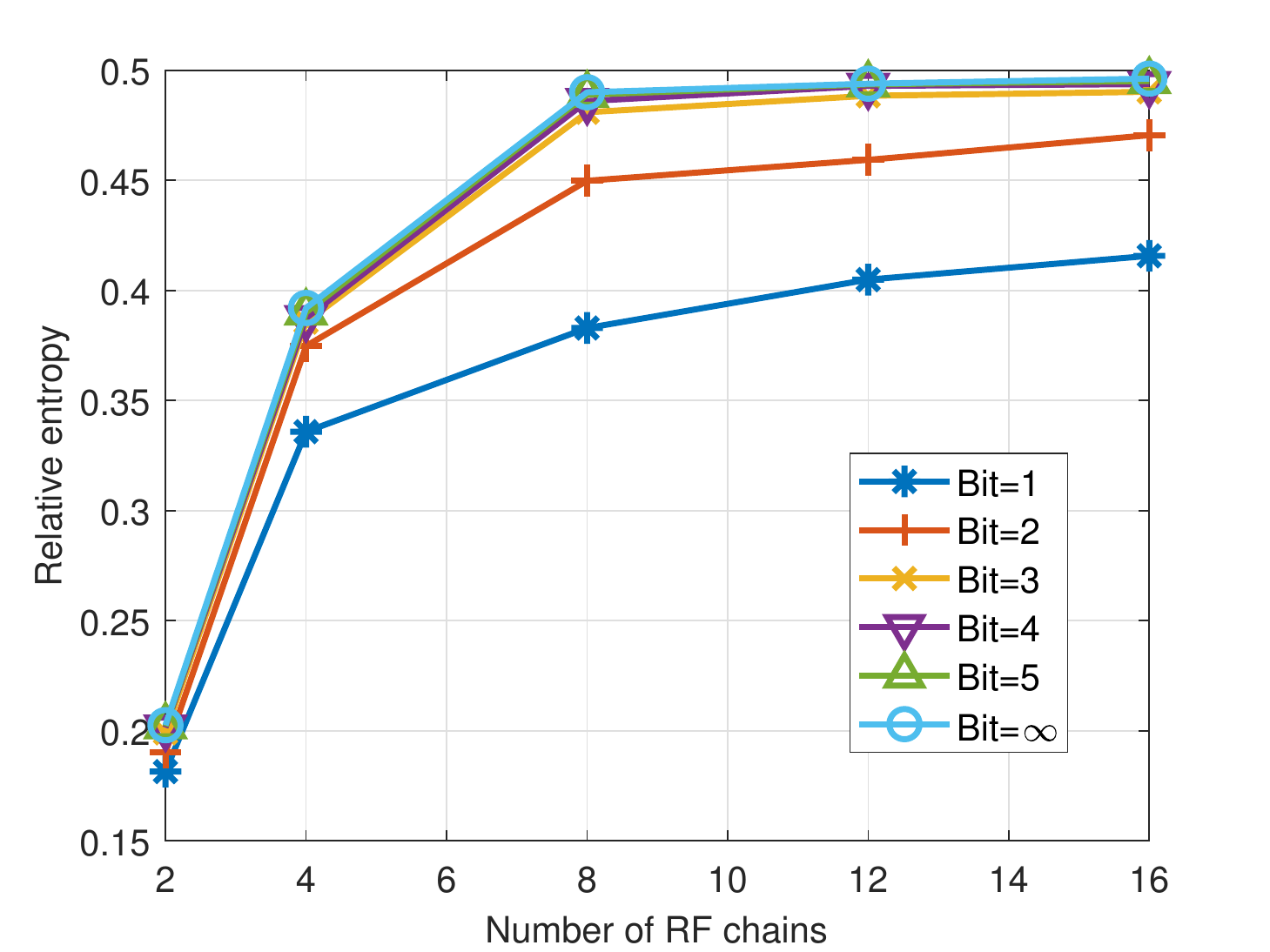}
	\caption{The relative entropy values  versus the number of RF chains for different numbers of ADC bits $\rm Bit$.}
	\label{fig:subfig:2a}
\end{figure}

\begin{figure}[!t]
	\centering
	\includegraphics[width=0.8\linewidth]{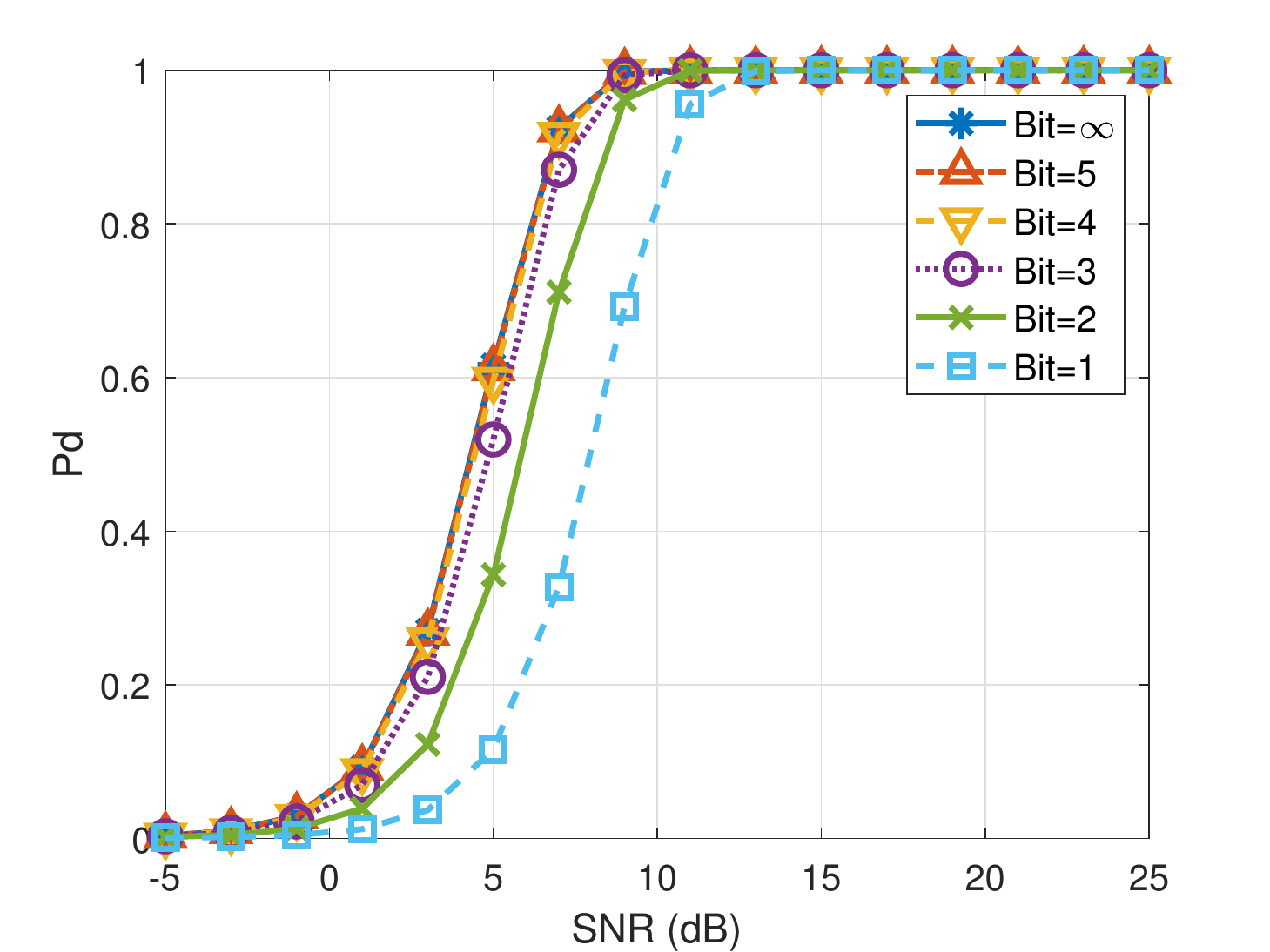}
	\caption{The detection performance  versus SNR for different  quantization levels $\rm Bit$. $ P_{\rm fa}=10^{-4} $.		\vspace{-2em}}
	\label{fig:subfig:2b}
\end{figure}

	{\textit{Example 2:}} In example 2, we have evaluated the  relative  entropy value  of the designed scheme for different numbers of ADC bits and RF chains. The remaining parameters are the same as example 1.  {Fig. \ref{fig:subfig:2a}   shows the relative  entropy value   versus the   number of RF {\color{black}chains} for different ADC bits. The results show that    the {\color{black}larger} the number of the RF chains, the better the relative  entropy can be achieved. This is because more RF chains mean  more {degrees of freedom (DoFs)} can be applied in the design stage. Moreover,  as the number of ADC bits   increase, the improvement of the  relative  entropy value  becomes  {\color{black}less and less apparent}. Specifically, when the ADC bit number is larger than 3,  the {\color{black}performance gap} with the ideal ADCs are very small. This suggests that there is no need  to adopt the high resolution ADCs in the receive system, and therefore,  the hardware cost can be    significantly reduced  in MIMO radar system. }
		The probabilities of detection
	associated with the designed beamformers for different ADC bits with $N_{\rm RF}=8$ is plotted in Fig. \ref{fig:subfig:2b}, where the probability of false alarm $ P_{\rm fa}=10^{-4} $, the number of Monte-Carlo trails is $10^6$ and the SNR is defined as ${\rm SNR}= \sigma_t^2/\sigma_n^2$. The result shows that,   as the number of {ADC bits increase, and the} probability of detection will be better and better.

\begin{figure}[!t]
		\centering
		\includegraphics[width=0.8\linewidth]{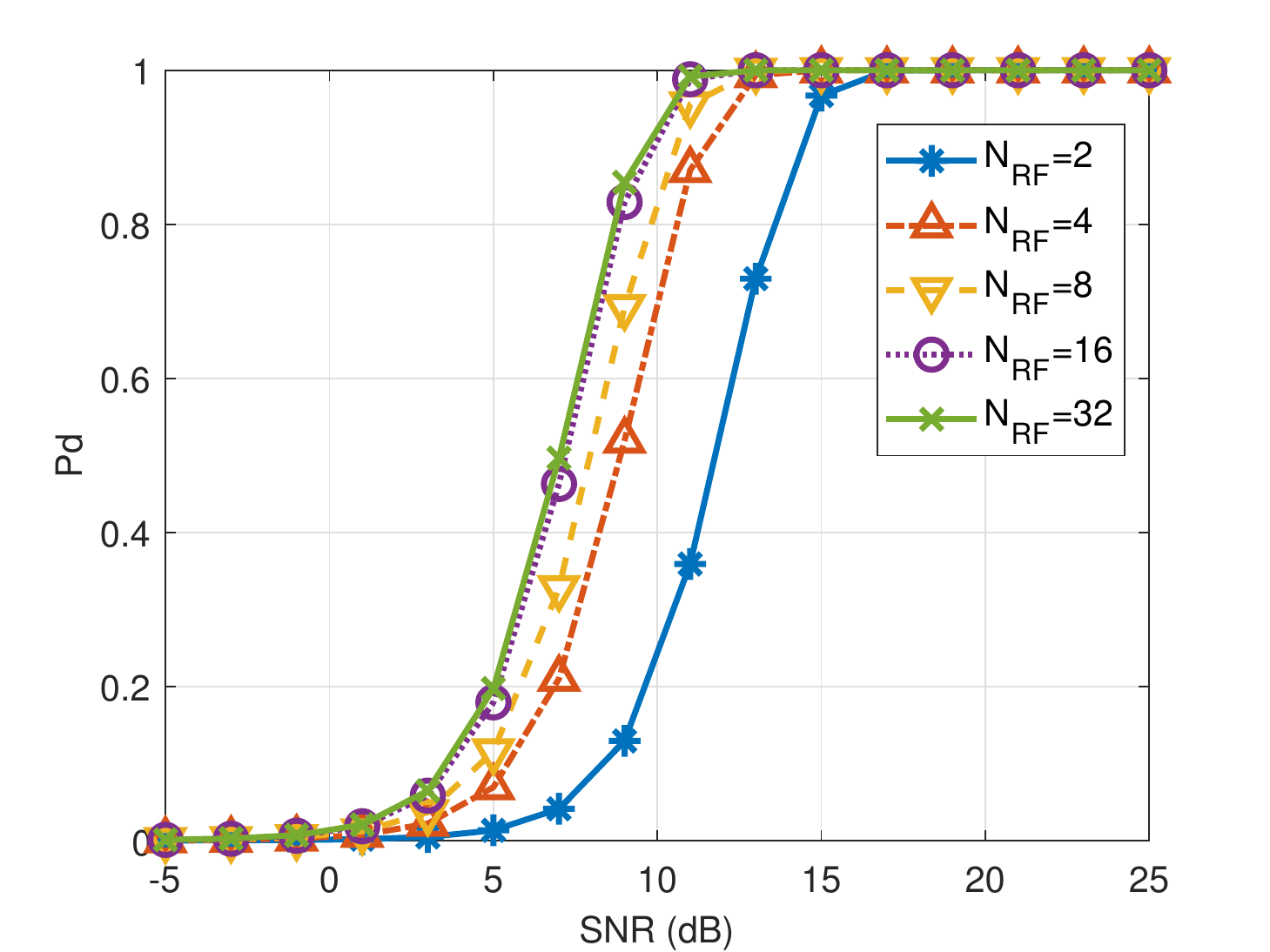}
		\caption{\color{black}The detection performance  versus SNR for different  numbers of the RF chains. $ P_{\rm fa}=10^{-4} $, ${\rm Bit}=1$.		\vspace{-1em}}
		\label{fig:subfig:3b}
	\end{figure}

	Fig. \ref{fig:subfig:3b}  displays the probabilities of detection of the designed scheme for different numbers of RF chains with one-bit ADCs,  this result shows that the probability of detection the beamformer with $ N_{\rm RF}=4 $ is obviously better than that of the beamformer with $ N_{\rm RF}=2 $,  {\color{black}but this  improvement becomes marginal when the $ N_{\rm RF} $  continues to increase.} Particularly, {\color{black}the beamformers with $ N_{\rm RF}=16 $ and $ N_{\rm RF}=32$ have almost identical performance.}
	
\begin{figure}[h]
  	\centering
  		\subfigure[]{
  		\includegraphics[width=0.8\linewidth]{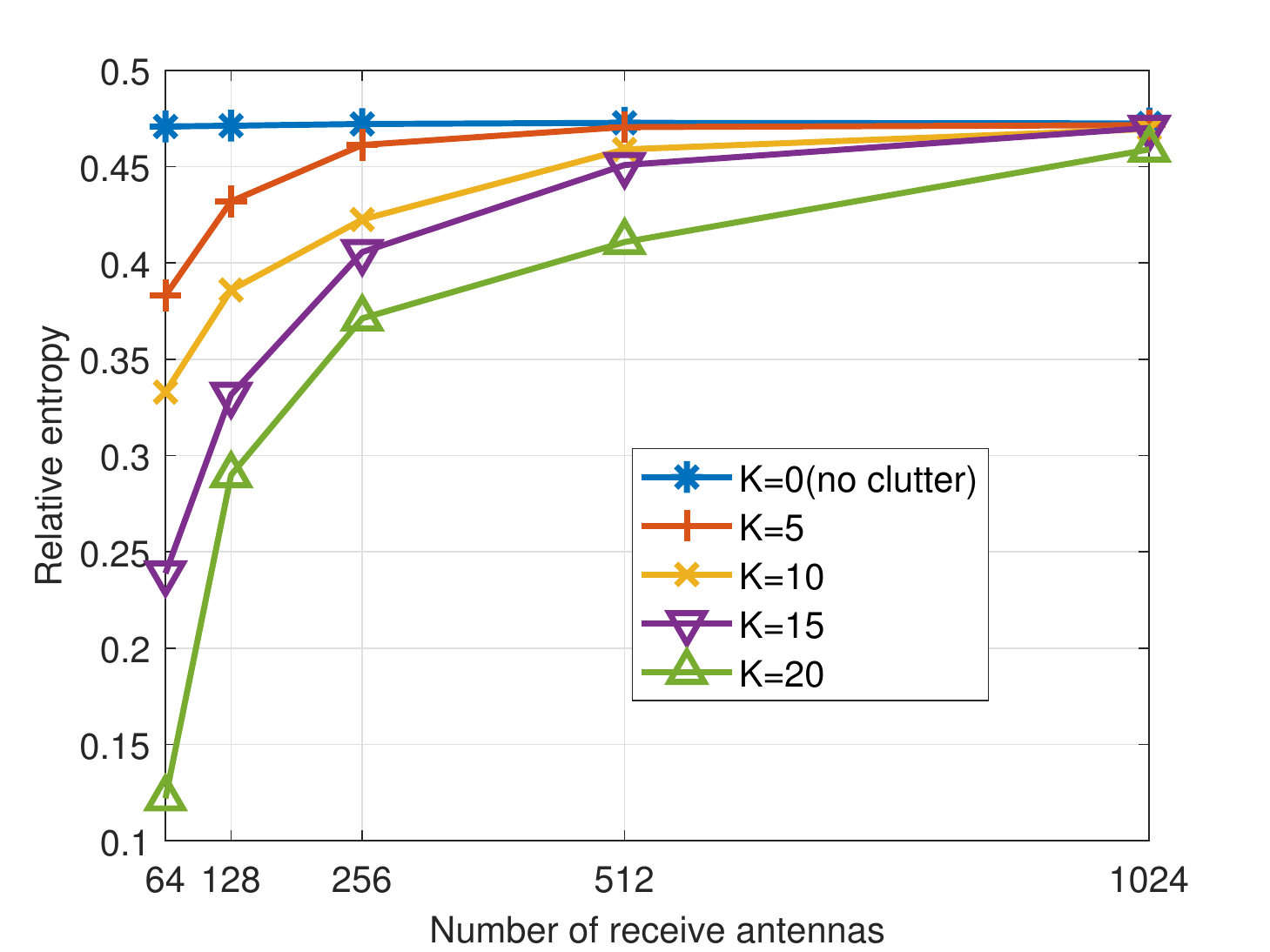}} 
  			\subfigure[]{
  			\includegraphics[width=0.8\linewidth]{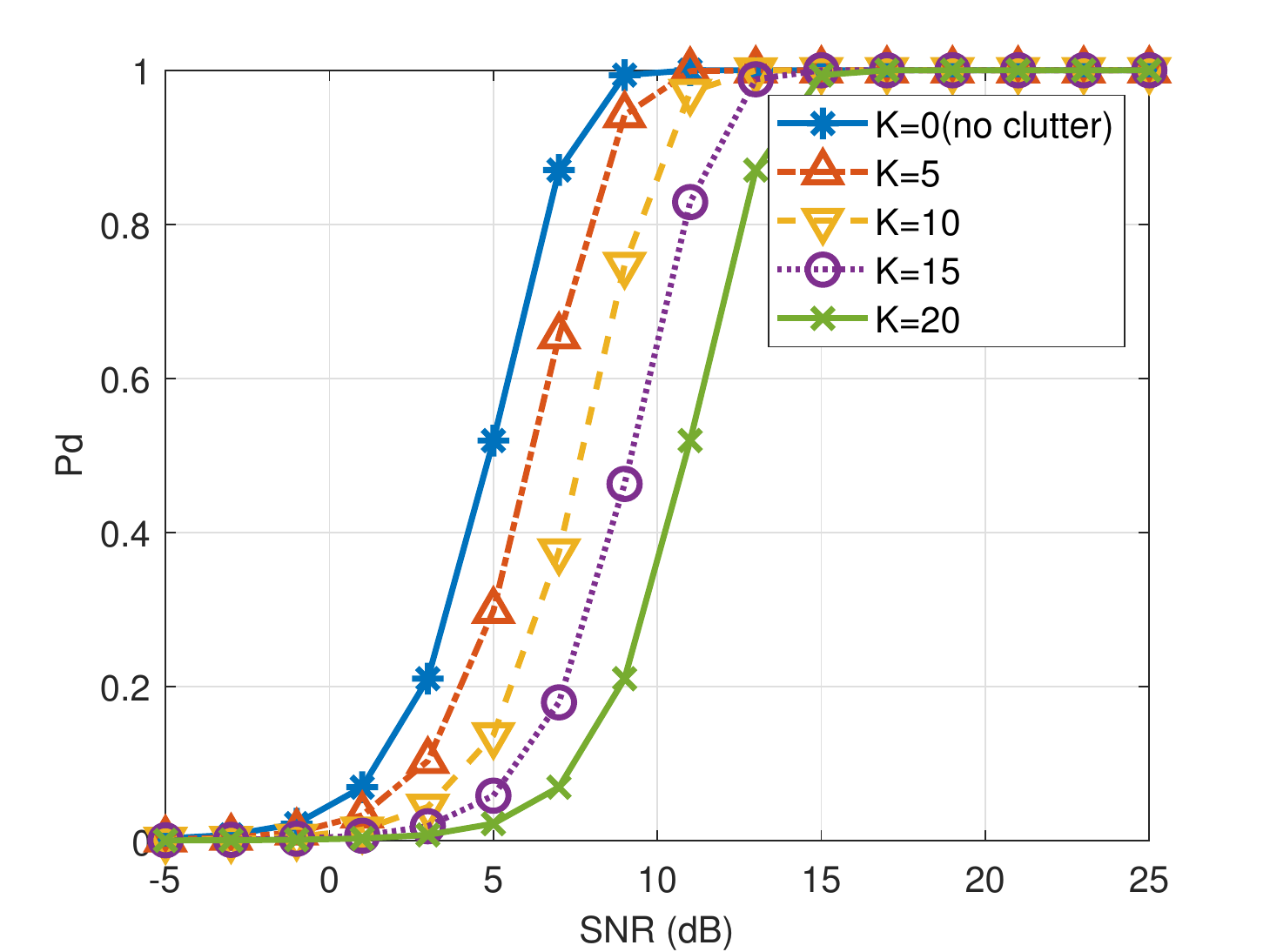}
  			}
  	\caption{Performance of the proposed design for {\color{black}a different number  of clutters}. (a) The relative
entropy values versus the number of receive antennas. (b) The detection performance versus SNR value for $P_{\rm fa}=10^{-4}$, $N_t=N_r=128$. }
  	\label{fig:pic5}
  \end{figure}

\begin{figure}[!t]
		\centering
		\includegraphics[width=0.8\linewidth]{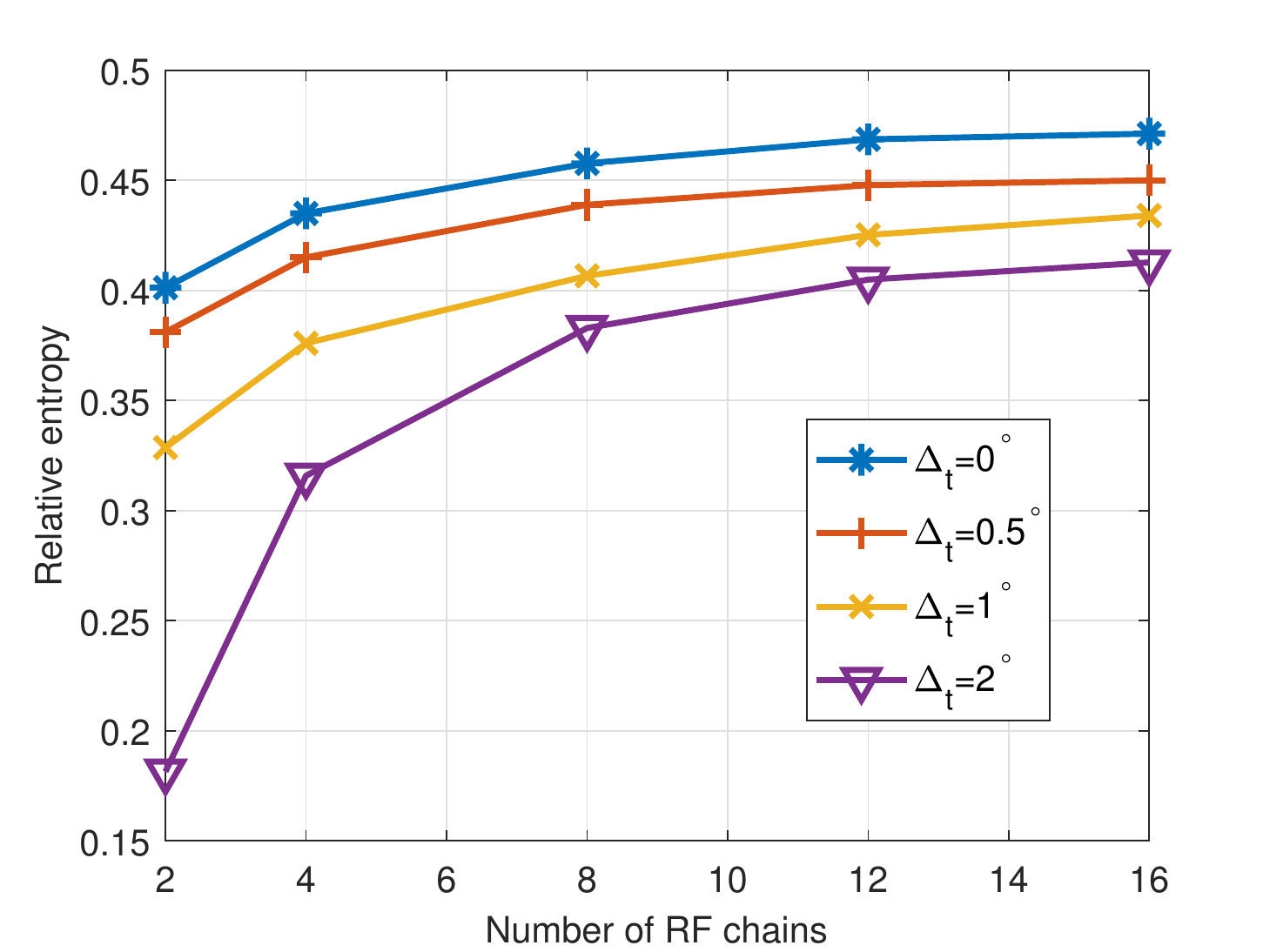}
		\caption{The relative entropy values  versus the number of RF chains when considering  different  uncertainties on $\theta_t$.}
		\label{fig:pic4}
	\end{figure}

{	{\textit{Example 3:}} In this example,  the effect of the number of clutters on the performance of the proposed method is assessed in Fig. \ref{fig:pic4}. To be specific, we consider the five cases, i.e., $K=0$ (no clutter), $K=5$, $K=10$, $K=15$, $K=20$. The specific directions of clutter of the considered case are listed in Tab. 1. The remaining parameters are the  same as example 1. 
   Fig. \ref{fig:pic5}  plots the relative  entropy versus    numbers of receive antennas $N_r$ {\color{black}a different number  of clutters}  $ K $. Note that the curve of $K=0$ provides the upper bound for our design.   The result  shows that  for a fixed $N_r$, the performance degradation increases with the increasing of the number of clutters increases. Moreover, we find that as the number of clutters decreases, the improvement provided by adding the number of receive antennas $N_r$  becomes more and more marginal. Particularly, when $K=0$, {\color{black}the relative entropy value does not decrease with increasing $N_r$}.  
   The reason lies in that the asymptotic error $ {\bf A}_{tc}^H {\bf A}_{tc} \to {\bf I}_{K+1} $ in Lemma 1 becomes smaller as the number of receive antennas increases, and the performance loss caused by our model is smaller. 
  Fig. \ref{fig:pic5}(b) shows the detection performance versus SNR value for $K=10$, $N_{\rm RF}=8$ and $P_{\rm fa}=10^{-4}$ when considering $N_t=N_r=128$. The result agrees with our expectation. 
 }

	{\textit{Example 4:}}	Fig. \ref{fig:pic4} shows the   relative  entropy as a function of the   number of RF chains for different  uncertainties on $\theta_t$, i.e., $\Delta_t$, when considering $K=10$.   It is {not surprising} to find  that the larger inaccuracies
	in the knowledge of  $\theta_t$, the greater
	the   relative  entropy loss will be.  Additionally, we also observe that as the number of RF chains increases, {\color{black}the relative  entropy  values become larger and larger.} 


	\subsection{Transmit beamforming with  one-bit phase shifters}
	In this subsection, we  assess  the performance  of the proposed hybrid beamforming design  for  the case of one-bit phase shifters adopted at the transmitter.
	
\begin{figure}[!t]
		\centering
		\subfigure[]{
			\label{fig:subfig:5a} 
			\includegraphics[width=0.8\linewidth]{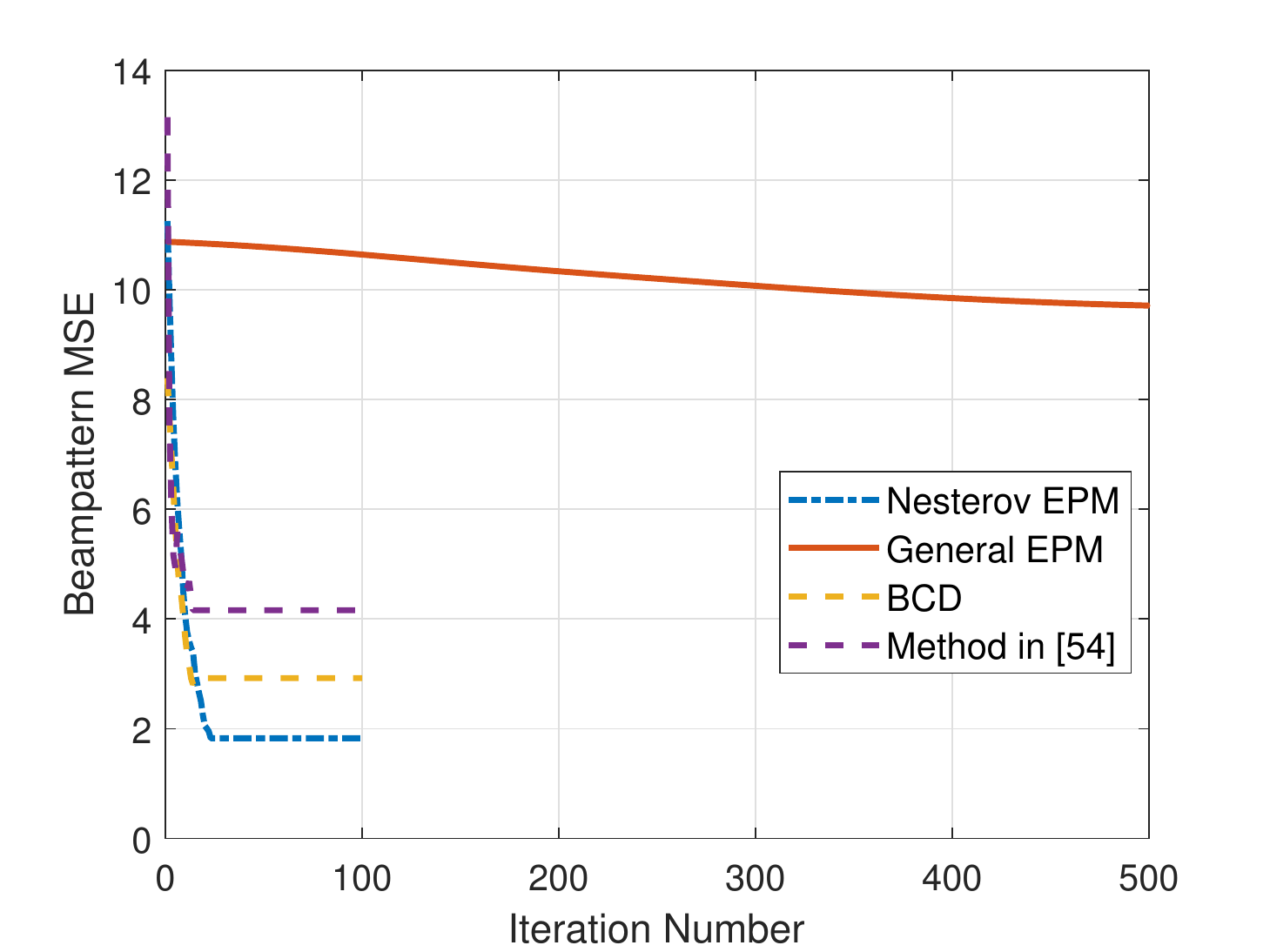}}
		\hspace{-0.2in}
		\subfigure[]{
			\label{fig:subfig:5b} 
			\includegraphics[width=0.8\linewidth]{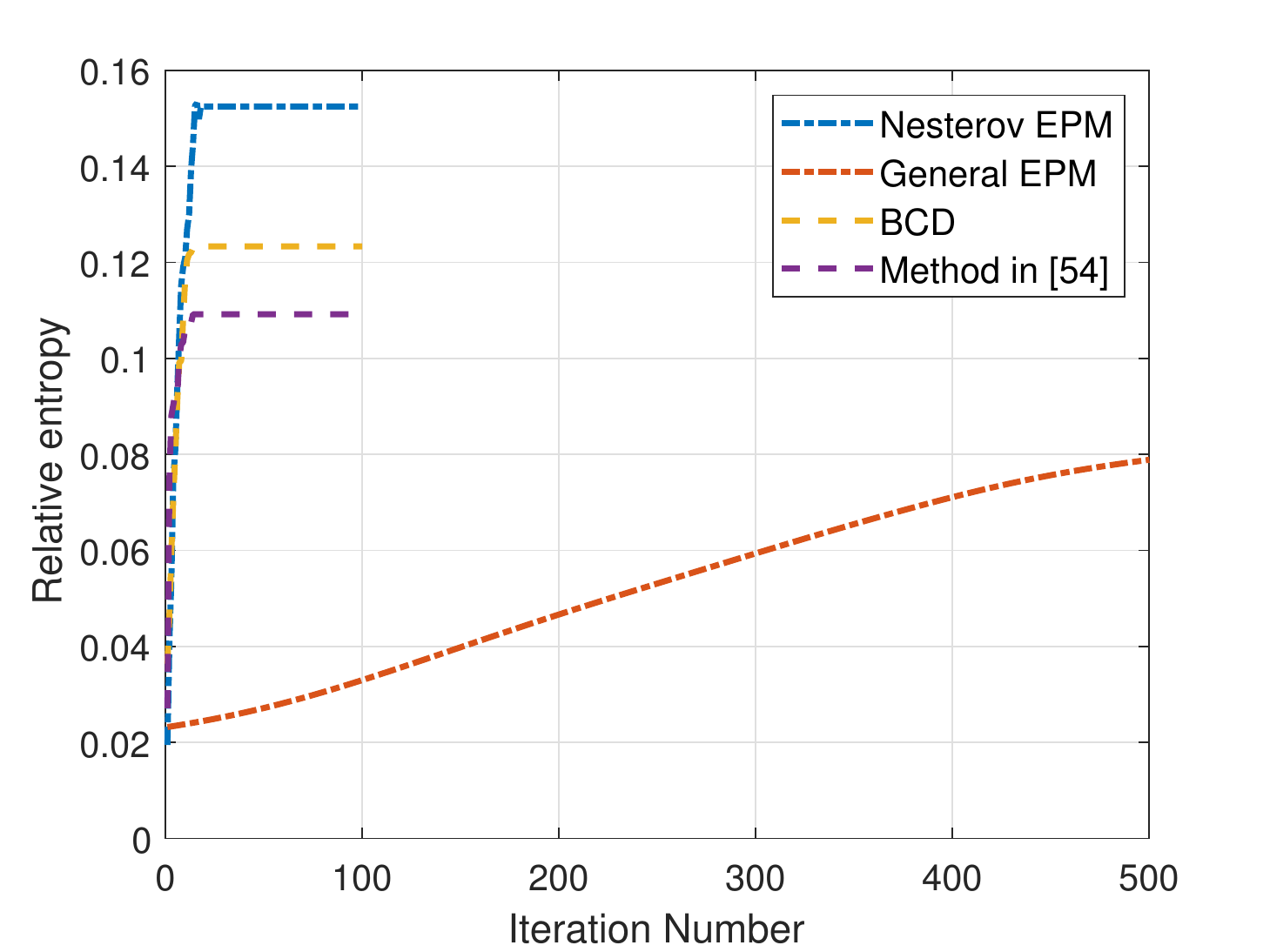}}
		\caption{The convergence performance of the proposed one-bit beamforming method. (a) The  beampattern MSE of the designed one-bit beamformers  versus iteration number, (b) the relative entropy of the designed one-bit beamformers  versus iteration number.}
		\label{fig:pic55} 
	\end{figure}

		{\textit{Example 5:}} The beampattern MSE property  and relative  entropy   
	of  one-bit beamformer  by using the EPM  with Nesterov-like gradient method ({\color{black}denoted  by} ``Nesterov EPM")  are plotted in Fig.  \ref{fig:pic55}, where we set $N_t=N_r=128$, $N_{\rm RF}=8$, and assume one-bit ADCs adopted at the receiver, {We consider   $10$ clutter scatterers  listed in Tab. I.}. Moreover, the BCD method, the EPM method with general gradient method  ({\color{black}denoted  by} ``general EPM")  and { the two-stage method in \cite{8359370} }    are also considered for comparison.   Fig.  \ref{fig:pic55}(a) shows the beampattern MSE values  of the designed beamformers. From the figure, we note that the proposed ``Nesterov EPM"   achieves a better MSE performance than the BCD, the gap is about $ 0.9  $. {\color{black}Besides, it is found that the ``Nesterov EPM" has faster convergence compared to the ``general EPM". The reason for the poor performance of  "General EPM” is that it is based on the traditional gradient method, which has a very slow convergence rate comparing to the Nesterov-like gradient method.} Fig.  \ref{fig:pic55}(b) shows the relative  entropies of the  designed beamformers with the three methods.  The result is consistence with that in Fig.  \ref{fig:pic55}(a).    

\begin{figure}[!t]
		\centering
		\subfigure[]{
			\label{fig:subfig:6a} 
			\includegraphics[width=0.8\linewidth]{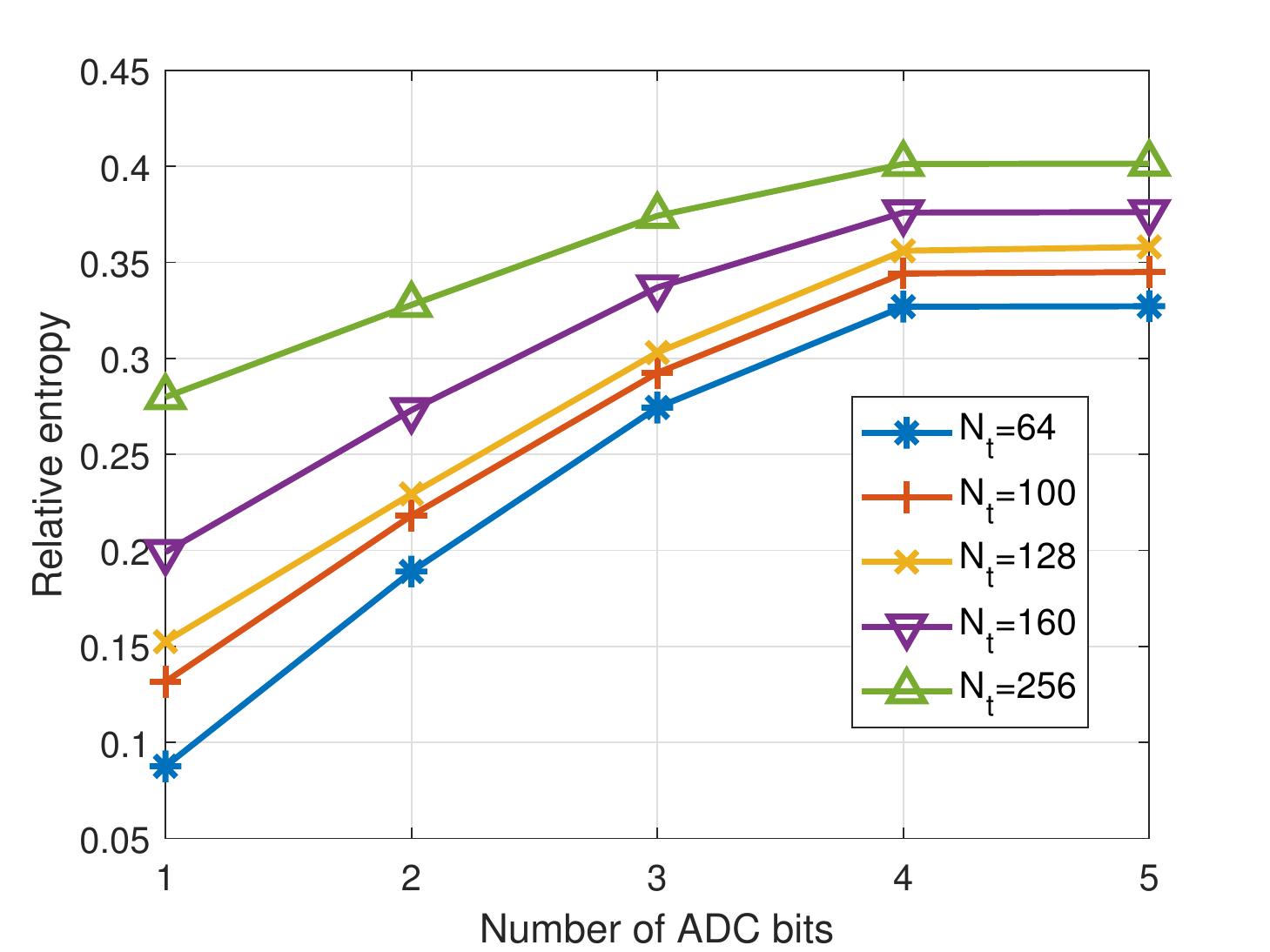}}
		\hspace{-0.2in}
		\subfigure[]{
			\label{fig:subfig:6b} 
			\includegraphics[width=0.8\linewidth]{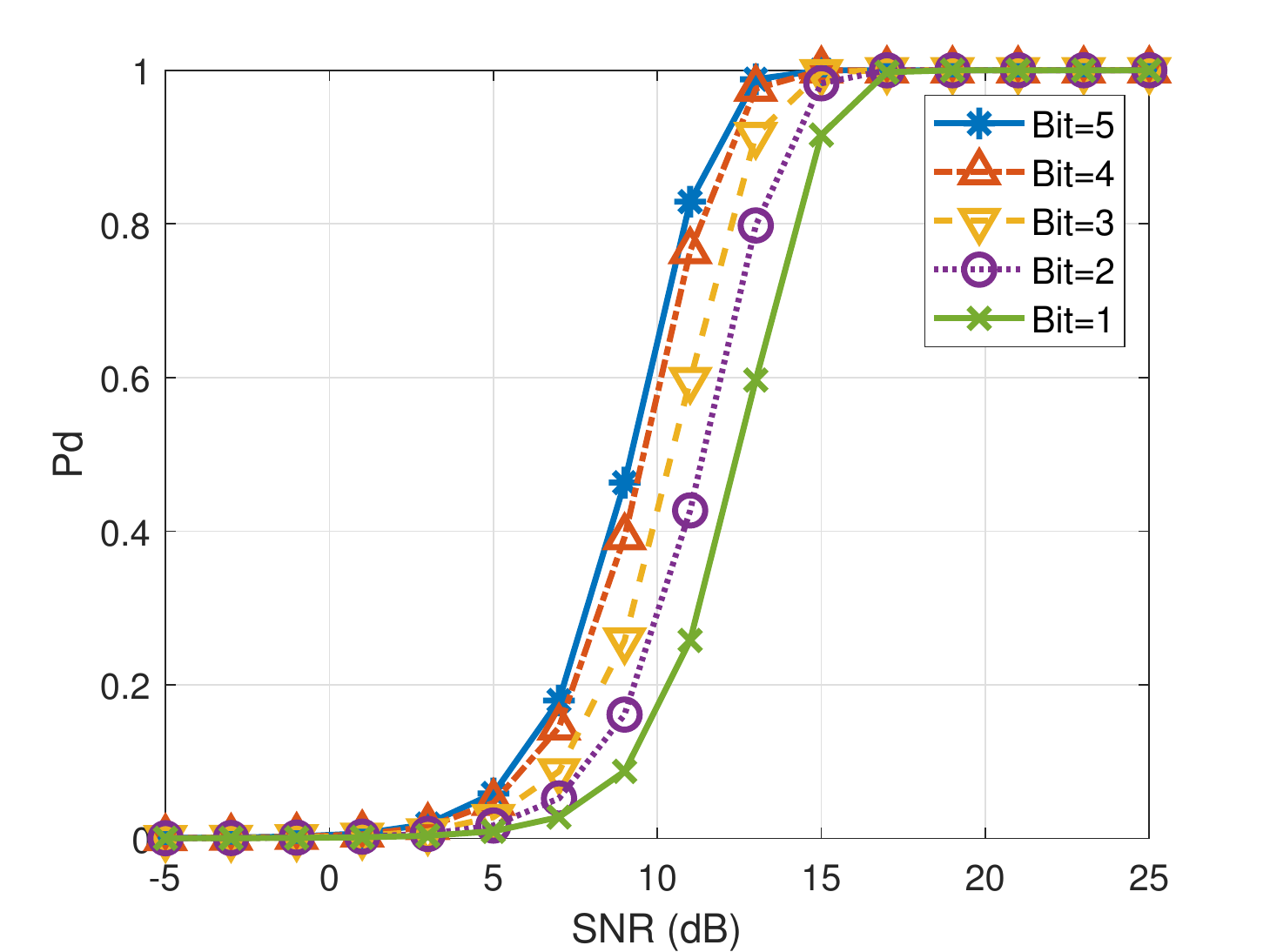}}
			\subfigure[]{
			\label{fig:subfig:6c} 
			\includegraphics[width=0.8\linewidth]{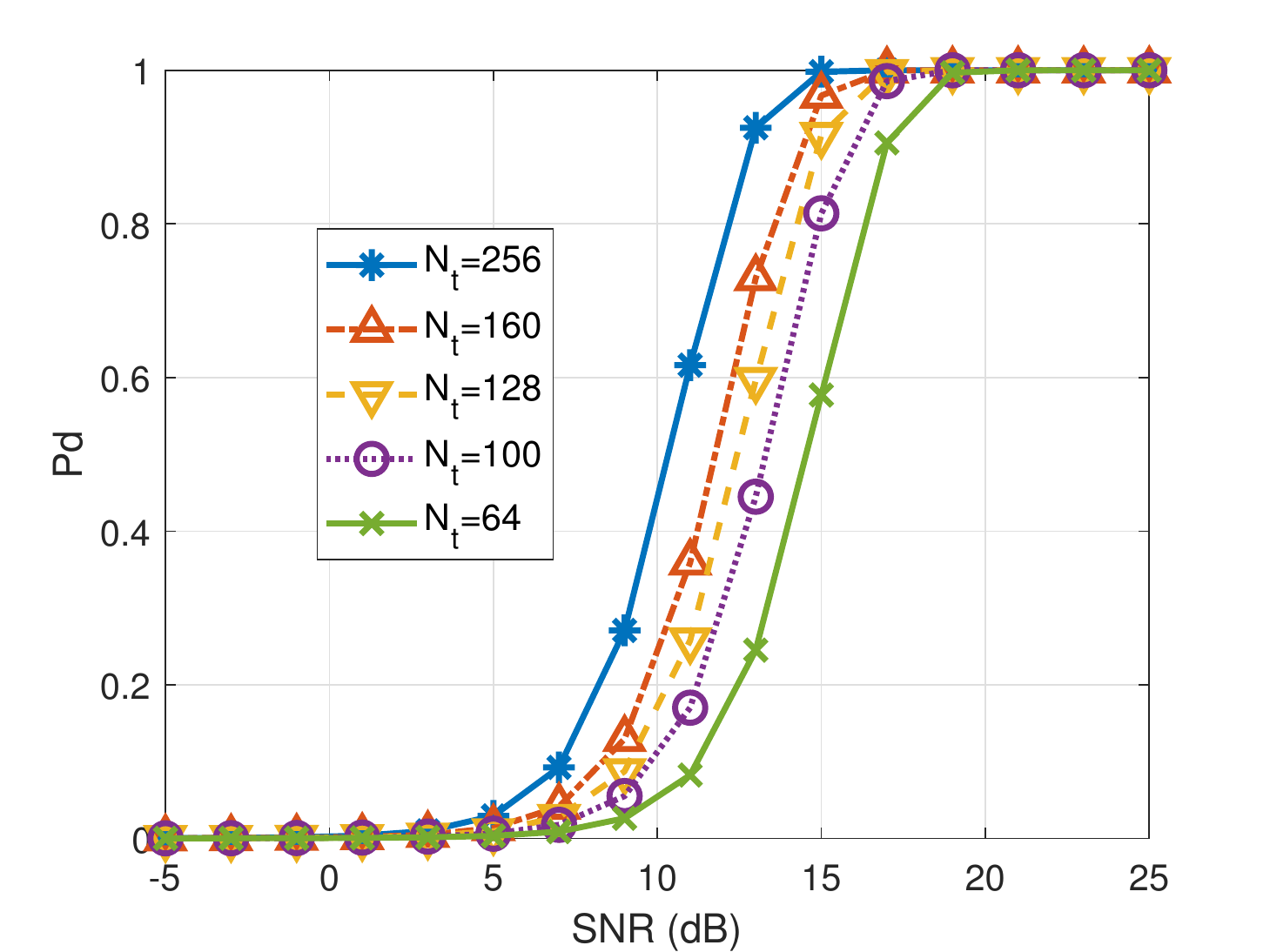}}
		\caption{Performance of the designed one-bit  beamformers. (a) The relative entropy values  versus number of ADC bits for different numbers of transmit antennas, (b) The detection performance  versus SNR for different  quantization levels $\rm Bit$, $N_t=128$.  $ P_{\rm fa}=10^{-4} $. (c) The detection performance  versus SNR for different  numbers of  transmit antennas when considering one-bit ADCs adopted at the receiver. $ P_{\rm fa}=10^{-4} $ }
		\label{fig:pic6} 
	\end{figure}

	Fig. \ref{fig:pic6}(a)  
	shows  the relative entropy values  of one-bit beamformer versus the number of  ADC bits for different numbers of transmit antennas.  Apparently, as the number of the ADC bit  increases, the relative entropy of one-bit beamformer becomes better and better, and when  the ADC bit is large than 4, there is no significant improvement of the relative entropy
	by increasing the  ADC bit.  This clearly
	{motivates us to adopt} the low-resolution ADCs at the receiver,
	since it is favored by reducing the hardware cost and circuit power consumption. Besides,  from the figure, we see that the system with one-bit phase shifters and one-bit ADCs benefits a lot from increasing the number of the transmit antennas, which suggests that we can  increase the number of antennas to reduce the loss caused by  one-bit phase shifters.    Fig.\ref{fig:pic6}(b) and Fig.\ref{fig:pic6}(c) plot   the  probabilities of detection versus SNR value for different numbers of ADC bits and transmit antennas, respectively. The results agree with our expectations. 

	\section{Conclusion}
	This paper has considered the problem of   constant-envelope beamforming design for mmWave large-scale system with low-cost hardware architecture, in which the transmit array adopts a hybrid digital/analog beamforming and the receive array adopts  low-resolution ADCs. 
	The corresponding problem of constant-envelope beamforming  is formulated via the relative entropy criterion.  To deal with {\color{black}the  nonconvex   problem}, we have developed  a two-stage  optimization framework based on  the MM technique. We also have considered the constant-envelope beamforming 
	 with  one-bit phase shifters, and developed an efficient iteration method based on the EPM and Nesterov-like method.
	The performance 
	of the proposed  schemes in terms of the relative entropy value and  {\color{black}detection performance} are assessed by numerical simulations.  The simulation results show  that the proposed  beamforming  is more efficient 
	than {\color{black}the existing methods}.  More importantly, we also find that  as the number of ADC bits   increases, the improvement of the detection performance becomes  more and more inapparent. Specifically, when the ADC bit number is larger than 3,  the {performance gap}  to the ideal ADCs is very small. In addition, when considering one-bit beamforming, we observe that the system benefits a lot from increasing the number of the transmit antennas.

	\appendices

		\section{Proof of  Lemma \ref{lem:1}}\label{prf:lem1}
	{

Considering that the $K+1$ { independent directions $ \{ \theta_k\}$ are  randomly }and uniformly distributed in $ [-\pi/2, \pi/2] $,  we have 
		\begin{equation} 
		\setlength{\abovedisplayskip}{3pt}
		    	{\bf a}_{r}^H(\theta_k) {\bf a}_{r}(\theta_l)  = \frac{1}{{{N_r}}}\sum\limits_{n = 0}^{{N_r} - 1} {{e^{j\pi n(\sin \theta_k-\sin \theta_l)}}} 
		\end{equation}
		
   Then, we get
	\begin{equation}
		\begin{aligned}
		& {\mathbb E}\left(  {\bf a}_{r}^H(\theta_k) {\bf a}_{r}(\theta_l) \right)  =\frac{1}{{{N_r}}}+\frac{1}{{{N_r}}}\sum\limits_{n = 1}^{{N_r} - 1} 	{\mathbb E}\left( {{e^{j\pi n (\sin {\theta_k-\sin \theta_l}) }}} \right)  \\
		&\quad =\frac{1}{{{N_r}}}+\frac{1}{{{N_r}}}\sum\limits_{n = 1}^{{N_r} - 1} 	{\mathbb E}_{\theta_l}\left( {{e^{-j\pi n  \sin \theta_l  }}} \right) {\mathbb E}_{\theta_k} \left( {{e^{ j\pi n  \sin  \theta_k  }}} \right)
		\end{aligned}
	\end{equation}
	
	Now, we calculate ${\mathbb E}_{\theta_k}\left( {{e^{j\pi n  \sin \theta_k  }}} \right)$ as 
	\begin{equation}
	    \begin{aligned}
	        {\mathbb E}_{\theta_k}\left( {{e^{j\pi n  \sin \theta_k  }}} \right)&=\frac{1}{\pi }\int_{ - \frac{\pi }{2}}^{\frac{\pi }{2}} {{e^{j\pi n\left( {\sin {\theta _k}} \right)}}{\rm{d}}{\theta _k}} \\
	        & = \frac{1}{\pi }\int_{ - \frac{\pi }{2}}^{\frac{\pi }{2}} {\cos (\pi n\sin {\theta _k}) + j\sin (\pi n\sin {\theta _k}){\rm{d}}{\theta _k}} 
	    \end{aligned}
	\end{equation}
Due to the fact that  $\cos(\pi n\sin {\theta _k})$ and  $\sin (\pi n\sin {\theta _k})$ are even and odd functions with respect to $\theta_k$, respectively, we have 
	\begin{equation}
	    \begin{aligned}
	        {\mathbb E}_{\theta_k}\left( {{e^{j\pi n  \sin \theta_k  }}} \right)&=\frac{2}{\pi }\int_0^{\frac{\pi }{2}} {\cos (\pi n\sin {\theta _k}){\rm{d}}{\theta _k}} \\
	        &= \frac{1}{\pi }\int_0^ {\pi }  {\cos (\pi n\sin {\theta _k}){\rm{d}}{\theta _k}} \\
	        &= \frac{1}{2\pi }\int_{0}^ {2\pi }  {\cos (\pi n\sin {\theta _k}){\rm{d}}{\theta _k}} \buildrel \wedge \over = 
	        {J}_0(\pi n)
	    \end{aligned}
	\end{equation}
	where ${J}_0(\pi n)$ is the Bessel function with the  zero-th  order \cite{abramowitz1964handbook}. Thus, we have 
	\begin{equation}
		\begin{aligned}
		& {\mathbb E}\left(  {\bf a}_{r}^H(\theta_k) {\bf a}_{r}(\theta_l) \right)  =
		\frac{1}{{{N_r}}}+\frac{1}{{{N_r}}}\sum\limits_{n = 1}^{{N_r} - 1} 	{J}_0^2(\pi n)
		\end{aligned}\label{aaa}
	\end{equation}
	According to the property of ${J}_0(\pi n)$ 
	that the  series $\{{J}_0(\pi n)\}_{n=1}^{N_r-1}$ is attenuated,  and ${J}_0(\pi n) \to \sqrt {\frac{2}{{\pi (\pi n)}}} \cos \left( {\pi n - \frac{\pi }{4}} \right) = {( - 1)^n}\sqrt {\frac{1}{{{\pi ^2}n}}} $ for large $n$ (\cite{abramowitz1964handbook}, section 9.2.1). Then, we have that
\begin{equation}  
\lim_{n \to \infty} \frac{ \frac{1}{N_r} J_0^2(\pi n) }{ \frac{1}{N_r \pi^2 n} } = 1
\end{equation}

Since $J_0^2(\pi n)>0, \frac{1}{  \pi^2 n} >0$, there must exist a constant $\cal K $ such that when $n>\cal K $, we have
	\begin{equation} 
 \frac{ \frac{1}{N_r} J_0^2(\pi n) }{ \frac{1}{N_r \pi^2 n} } <2
\end{equation}
Then, one gets 
\begin{equation}
\begin{aligned}
0&\le \frac{1}{N_r} \sum_{n=1}^{N_r} J_0^2(\pi n) 
= \frac{1}{N_r} \sum_{n=1}^{\cal K} J_0^2(\pi n) + \frac{1}{N_r} \sum_{ { {n=\mathcal{K}+1}} }^{N_r} J_0^2(\pi n)\\
&\le   \frac{1}{N_r} \sum_{n=1}^{\cal K} J_0^2(\pi n) + \frac{2}{N_r} \sum_{n={\mathcal{K}}+1}^{N_r} \frac{1}{  \pi^2 n} \\
&\le \frac{1}{N_r} \sum_{n=1}^{\cal K} J_0^2(\pi n) + \frac{2}{N_r} \sum_{n= 1}^{N_r} \frac{1}{  \pi^2 n} 
\end{aligned}\label{50a}
\end{equation}

Due to the fact that 	
	\begin{equation}   \label{eq:b}
0<\frac{1}{N_r \pi^2 n} < \frac{1}{N_r^{(1-\epsilon)} \pi^2 n^{(\epsilon+1)}}
\end{equation}
and that  
\begin{equation}
    \mathop {\lim }\limits_{{N_r} \to \infty } \frac{1}{{N_r^{(1 -\epsilon )}}}\sum\nolimits_{n = 1}^{{N_r}} {\frac{1}{{{\pi ^2}{n^{( \epsilon+ 1)}}}}} =0
\end{equation}
where $0<\epsilon<1$, we can obtain that 
\begin{equation}
    \mathop {\lim }\limits_{{N_r} \to \infty } \frac{1}{{N_r}}\sum\nolimits_{n = 1}^{{N_r}} {\frac{1}{{{\pi ^2}{n}}}} =0\label{53}
\end{equation}

On the other hand, since $\sum_{n=1}^{\cal K} J_0^2(\pi n)$ is bounded, one can arrive at 
\begin{equation}
    \mathop {\lim }\limits_{{N_r} \to \infty } \frac{1}{{N_r}}\sum_{n=1}^{\cal K} J_0^2(\pi n)=0
    \label{54}
\end{equation}

Substituting \eqref{54} and \eqref{53} into \eqref{50a}, one gets that 
\begin{equation}
     \begin{aligned}
0&\le   \mathop {\lim }\limits_{{N_r} \to \infty } \frac{1}{N_r} \sum_{n=1}^{N_r} J_0^2(\pi n) \\
&\le   \mathop {\lim }\limits_{{N_r} \to \infty } \frac{1}{N_r} \sum_{n=1}^{\cal K} J_0^2(\pi n) + \frac{2}{N_r} \sum_{n= 1}^{N_r} \frac{1}{  \pi^2 n}=0
\end{aligned}
\end{equation}
 Thus, $ \mathop {\lim }\limits_{{N_r} \to \infty } \frac{1}{N_r} \sum_{n=1}^{N_r} J_0^2(\pi n) =0$, which  implies that  $  	{\mathbb E}\left(  {\bf a}_{r}^H(\theta_k) {\bf a}_{r}(\theta_l) \right) \to 0  $ when $ N_r \to \infty  $.}
	
	Furthermore,   we have $ {\mathbb V}{\rm ar}\left(   {\bf a}_{r}^H(\theta_k) {\bf a}_{r}(\theta_l)  \right)  $ as  
 \begin{subequations}
	\begin{align}
	&	{\mathbb V}{\rm ar}\left(   {\bf a}_{r}^H(\theta_k) {\bf a}_{r}(\theta_l)  \right)  ={\mathbb E}\left( {{{\left| {{\bf{a}}_r^H({\theta _k}){{\bf{a}}_r}({\theta _l})} \right|}^2}} \right) \notag\\
	&  = \frac{{{N_r} - 1}}{{N_r^2}} + \frac{1}{{N_r^2}}\sum\limits_{m \ne n}^{{N_r} - 1} {\sum\limits_{n = 1}^{{N_r} - 1} {\mathbb E} \left( {{e^{j\pi (m - n)(\sin {\theta _k} - \sin {\theta _l})}}} \right)}  + \frac{1}{{N_r^2}} \notag\\
	&\qquad + \frac{2}{{N_r^2}}\Re \left\{ {\sum\limits_{n = 1}^{{N_r} - 1}  {\mathbb E}\left( {{e^{j\pi n\left( {\sin {\theta _k} - \sin {\theta _l}} \right)}}} \right)} \right\} \notag\\
	& 	  = \frac{1}{{N_r }}+\frac{1}{{N_r^2}}\sum\limits_{m \ne n}^{{N_r} - 1} {\sum\limits_{n = 1}^{{N_r} - 1} {\mathbb E} \left( {{e^{j\pi (m - n)(\sin {\theta _k} - \sin {\theta _l})}}} \right)} \notag\\
	&=\frac{1}{{N_r }}+\frac{1}{{N_r^2}}\sum\limits_{m \ne n}^{{N_r} - 1} {\sum\limits_{n = 1}^{{N_r} - 1}   {J}_0^2(\pi (m-n)) } \notag
	\end{align}
\end{subequations}
		
Further, we obtain that  
\begin{align}
\frac{1}{N_r^2} \sum_{m \neq n}^{N_r-1} & \sum_{n=1}^{N_r-1} J_0^2 (\pi(m-n)) \\
&\leqq  \frac{1}{N_r^2} \sum_{m=1}^{N_r-1} \sum_{n=1}^{N_r-1} J_0^2 (\pi(m-n))  \\
&\leqq  \frac{1}{N_r^2} \sum_{k=(2-N_r)}^{N_r-2}  (N_r-1) J_0^2 (\pi k)        \\
&{ \leqq \frac{1}{{{N_r}}}J_0^2(0) + \frac{2}{{N_r}}\sum\limits_{k = 1}^{{N_r} - 1} {J_0^2(\pi k)} } \\
& { = \frac{1}{{{N_r}}} + \frac{2}{{N_r}}\sum\limits_{k = 1}^{{N_r} - 1} {J_0^2(\pi k)} }
\end{align}

Thus, similar to Eq. \eqref{aaa}, we can infer   that $ 	{\mathbb V}{\rm ar}\left(   {\bf a}_{r}^H(\theta_k) {\bf a}_{r}(\theta_l)  \right)  \to 0$ when $ N_r \to \infty $.  

	\section{Proof of  Lemma \ref{lem:2}}\label{prf:lem2}
	Defining $\tilde{\bf X}={\bf X}{\bf X}^H$ and $ \tilde{\bf x}={\rm vec}(\tilde{\bf X}) $, we have 
	\[
	\begin{aligned}
	g({\bf X}) =\tilde{\bf x}^H({{\bf{u}}{{\bf{u}}^H} \otimes {{\bf{u}}^*}{{\bf{u}}^T}})\tilde{\bf x}      \buildrel \Delta \over =  g( \tilde{\bf x})\\
	\end{aligned}
	\]
	
	Using Taylor's theorem{\cite{8239862}},  the second-order inequality
	of $ g( \tilde{\bf x}) $ at  the point $  \tilde{\bf x}^{(m)} $ is given by
	\begin{equation}
	\begin{aligned}
	g( \tilde{\bf x}) &\le  g( \tilde{\bf x}^{(m)})\\
    & ~~~+ \Re\big({\tilde{\bf x}^{{(m)}H}}  ({{\bf{u}}{{\bf{u}}^H} \otimes {{\bf{u}}^*}{{\bf{u}}^T}})  (\tilde{\bf x}-\tilde{\bf x}^{(m)} )\big)\\
	& ~~~+\frac{{\lambda _{\rm max}\left( {{\bf{u}}{{\bf{u}}^H} \otimes {{\bf{u}}^*}{{\bf{u}}^T}} \right)}}{{\rm{2}}} \left\| \tilde{\bf x} -\tilde{\bf x}^{(m)}\right\|^2 \\
	& = \frac{{\lambda _{\rm max}\left( {{\bf{u}}{{\bf{u}}^H} \otimes {{\bf{u}}^*}{{\bf{u}}^T}} \right)}}{{\rm{2}}}  
	\left\|  \tilde{\bf X} -\tilde{\bf X}^{(m)}   \right\|_F^2  \\
	&~~~ + ( {{\bf{u}}^T}  {\tilde{\bf X}^{(m)}  }   {{\bf{u}}^*}  )^*    \left( {{\bf{u}}^T}  \tilde{\bf X}   {\bf{u}}^* \right)     
	&~~
	\end{aligned}\label{84}
	\end{equation}  
 Based on ${\bf T}^H{\bf T}  =  {\bf I}_{\rm RF} $ and  $\left| {{\bf{T}} }(i,j)\right|  = \frac{1}{\sqrt{N_t}}, \forall i, j$, one gets 
	\begin{equation}
	\begin{aligned}
	&  \left\| \tilde{\bf x} -\tilde{\bf x}^{(m)}\right\|^2 =\left\|  \tilde{\bf x}     \right\|^2+\left\|  \tilde{\bf x}^{(m)}    \right\|^2 -2\Re\big(   \tilde{\bf x}^H \tilde{\bf x}^{(m)}  \big)\\
	&  =  {\rm vec}^H({\bf X}{\bf X}^H) {\rm vec}({\bf X}{\bf X}^H) \\
	& ~~~+   {\rm vec}^H({\bf X}^{(m)}{\bf X}^{{(m)}H})  {\rm vec}({\bf X}^{(m)}{\bf X}^{{(m)}H}) \\
    & ~~~-2\Re\big( {\rm vec}^H({\bf X}{\bf X}^H) {\rm vec}({\bf X}^{(m)}{\bf X}^{{(m)}H}) \big)\\
	& ={\bf x}^H \big(  {\bf X}^T {\bf X}^* \otimes {\bf I}_{N_t}  \big){\bf x} +{\bf x}^{{(m)}H} \big(  {\bf X}^{{(m)}T} {\bf X}^{{(m)}*} \otimes {\bf I}_{N_t}  \big){\bf x}^{(m)}\\
	&~~~-2\Re\big({\rm Tr} ( {\bf X}{\bf X}^H {\bf X}^{{(m)}} {\bf X}^{{(m)}H}) \big)\\
	&=2N_{\rm RF} -2\Re\big({\rm Tr} ( {\bf X}{\bf X}^H {\bf X}^{{(m)}} {\bf X}^{{(m)}H}) \big)
	\end{aligned}\label{85}
	\end{equation}
	where $ {\bf x}= {\rm vec} \left({\bf X} \right)  $.
	
	Substituting \eqref{85} into \eqref{84}, and ignoring the constant terms unrelated to $\bf X$   yields 
	\begin{equation}
	\begin{aligned}
	& g(  {\bf X}) \le   ( {{\bf{u}}^T}  { {\bf X}^{(m)} {\bf X}^{{(m)}H} }   {{\bf{u}}^*}  )    \left( {{\bf{u}}^T}   {\bf X}  {\bf X}^H  {\bf{u}}^* \right)  \\
	& \quad -  {{{\color{black}\lambda _{\rm max}}\left( {{\bf{u}}{{\bf{u}}^H} \otimes {{\bf{u}}^*}{{\bf{u}}^T}} \right)}}   \Re\big({\rm Tr} ( {\bf X}{\bf X}^H {\bf X}^{(m)} {\bf X}^{{(m)}H}) \big) \\
		& \quad+{\color{black}N_{\rm RF}{{{\color{black}\lambda _{\rm max}}\left( {{\bf{u}}{{\bf{u}}^H} \otimes {{\bf{u}}^*}{{\bf{u}}^T}} \right)}}}
	\end{aligned}
	\end{equation}
	
	Thus, we complete this proof. 

    \section{Proof of Lemma \ref{lem:3}}\label{prf:lem3}
    
        According to Lemma \ref{lem:2} and \eqref{eq:33}, the majorizing function $\widetilde{\mathcal{Z}}(\mathbf{t},\mathbf{t}^{(m)})$ of $\mathcal{Z}(\mathbf{t})$ at the point $\mathbf{t}^{(m)}$ is defined as
        \begin{equation}
            \widetilde{\mathcal{Z}}(\mathbf{t},\mathbf{t}^{(m)}) = \Re ( \mathbf{t}^{(m)H} \widetilde{\mathbf{Q}}^{(m)} \mathbf{t})  + \text{const.} 
        \end{equation}
        where $\widetilde{\mathbf{Q}}^{(m)} = {\bf I}_{N_{\rm RF}}\otimes {\bf Q}^{(m)} - {\lambda _{\rm max}} ({\bf I}_{N_{\rm RF}}\otimes {\bf Q}^{(m)}  ){\bf I}$ and $\text{const} = \sum\limits_{\theta_p \in \Theta_P} {| \hat{\phi}(\theta_p)  |^2} + 2\varsigma N_{\rm RF} + N_{\rm RF}\lambda _{\rm max}  ({\bf I}_{N_{\rm RF}}\otimes {\bf Q}^{(m)}  )$.

        Based on the basis of MM \cite{7547360,6601713}, the majorizing function $\widetilde{\mathcal{Z}}(\mathbf{t},\mathbf{t}^{(m)})$ of $\mathcal{Z}(\mathbf{t})$ satisfy the following two conditions:
        \begin{subequations}
            \begin{align}
                & \widetilde{\mathcal{Z}}(\mathbf{t},\mathbf{t}^{(m)}) \ge \mathcal{Z}(\mathbf{t}) \label{prf:lem3-1a}\\
                & \widetilde{\mathcal{Z}}(\mathbf{t}^{(m)},\mathbf{t}^{(m)}) = \mathcal{Z}(\mathbf{t}^{(m)}) \label{prf:lem3-1b}
            \end{align}
        \end{subequations}

        Then, it is easy to show that with MM scheme, the objective value is monotonically decreasing at each iteration, i.e.,
        \begin{equation}
            \begin{aligned}
                \mathcal{Z}(\mathbf{t}^{(m+1)}) \mathop  \le \limits^{\text{(a)}} \widetilde{\mathcal{Z}}(\mathbf{t}^{(m+1)},\mathbf{t}^{(m)}) \mathop  \le \limits^{\text{(b)}}  \widetilde{\mathcal{Z}}(\mathbf{t}^{(m)},\mathbf{t}^{(m)}) \mathop  = \limits^{\text{(c)}}  \mathcal{Z}(\mathbf{t}^{(m)})
            \end{aligned}
        \end{equation}
        where (a) and (c) hold since the properties of the majorization function, namely \eqref{prf:lem3-1a} and \eqref{prf:lem3-1b} respectively.
        The inequality (b) hold since the solution to problem \eqref{35} is optimal.

        Thus, we complete this proof.

	\section{Proof of  Proposition \ref{pro:1}}\label{prf:pro1}
	According to the definition
	of $\Psi$ and  Cauchy-Schwarz inequality, we have that 
	\begin{equation}
	\frac{N}{M} = {\bf x}^T{\bf y} \le \|{\bf x}\| \|{\bf y}\| \le \|{\bf x}\| \sqrt{\frac{N}{M}}
	\end{equation}
	Thus, we have $\|{\bf x}\|\ge \sqrt{\frac{N}{M}} $. Combining
	the  set $ -\frac{1}{\sqrt{M}} \le {\bf x}\le \frac{1}{\sqrt{M}} $, one gets the following set 
	\[
	\|{\bf x}\|\ge \sqrt{\frac{N}{M}}, ~  0  \le |{x_n}|\le \frac{1}{\sqrt{M}}, \forall n
	\]
	Thus, we have $ |{x_n}|= \frac{1}{\sqrt{M}} $, i.e., $ \left\lbrace  {\bf x} \in \frac{1}{\sqrt{M}}  \left\{ -1, ~1 \right\}^{N} \right\rbrace   $.

	\section{COMPUTATION of   $ \nabla_{\bf t} {\cal F}_{\rho}({\bf t} ) $ }\label{prf:grad}
	{
	Let $ {\cal F}_{\rho}({\bf t} ) =\psi_1({\bf t})+ \varsigma \psi_2({\bf T})  $ denote the objective value of problem \eqref{29}, where $ \psi_1({\bf t}) =\sum_{\theta_p \in \Theta_P}  {( {\bf t}^T {\bf \Phi}_p{\bf t}    -\hat{\phi}_p )^2}+\rho(N_{\rm RF}- \sqrt{N_{\rm RF}} \|{\bf t} \|) $ and $  \psi_2({\bf T})  =  \left\|  {\bf T}^T{\bf T}-{\bf I}_{N_{\rm RF}} \right\|^2_F$. Then, we can obtain the derivation of  $ \psi_1({\bf t})  $ with respect to $t_{(j-1)N_t+i}, j=1,\cdots,N_{\rm RF}; i=1,\cdots,N_t$ as 
	\begin{equation}
		\begin{aligned}
	&\frac{{\partial {\psi _1}({\bf{t}})}}{{\partial {t_{(j - 1){N_t} + i}}}}  =\\
	&~~ \sum\limits_{\theta_p \in \Theta_P} {({{\bf{t}}^T}{{\bf{\Phi }}_p}{\bf{t}} - {{\hat \phi }_p}){{\bf{\Phi }}_{p,((j - 1){N_t} + i):}}{\bf{t}}}  - \frac{\rho \sqrt{N_{\rm RF}}}{\|{\bf t}\| }t_{(j - 1){N_t} + i}
		\end{aligned}
	\end{equation}
	where $  {{{\bf{\Phi }}_{p,((j - 1){N_t} + i):}}}  $ denotes the $(j - 1){N_t} + i  $-th row of the matrix $ {\bf{\Phi }}_p$ and $  t_{(j - 1){N_t} + i}  $ stands for the  $(j - 1){N_t} + i  $-th entry of the vector $ {\bf t}  $.
	
	To proceed, to {compute the derivative of}  $ \psi_2({\bf T})  $ with respect to $t_{(j-1)N_t+i} $ (i.e., $ T_{i,j} $), we need to extract the contribution of $ T_{ij} $ to  $ \psi_2({\bf T})  $. More exactly, let $  {\overline {\bf{T}} }_{i,j} $ be the matrix $\bf{T}  $  whose $ (i, j) $-th entry is
	zeroed, and $ {\bf E}_{i,j} $ be an $ N_t\times N_{\rm RF} $-dimensional matrix whose $ (i, j) $-th element is 1 and 0 otherwise, then we have 
	\begin{equation}
	\begin{aligned}
	\psi_2( T_{ij}) &=\left\| {{{\left( {{T_{ij}}{{\bf{E}}_{i,j}} + {{\overline {\bf{T}} }_{i,j}}} \right)}^T}\left( {{T_{ij}}{{\bf{E}}_{i,j}} + {{\overline {\bf{T}} }_{i,j}}} \right) - {{\bf{I}}_{{N_{{\rm{RF}}}}}}} \right\|_F^2\\
	&= {T_{ij}^4}+2T_{ij}^3{\rm Tr}\Big({\bf{E}}_{i,j}^T{\bf{E}}_{i,j}\big( {\bf{E}}_{i,j}^T {\overline {\bf{T}} }_{i,j}+ {\overline {\bf{T}} }_{i,j}^T  {\bf{E}}_{i,j}     \big)  \Big)\\
	&~~~+2T_{ij}^2 {\rm Tr}\Big({\bf{E}}_{i,j}^T{\bf{E}}_{i,j}\big(   {\overline {\bf{T}} }_{i,j}^T  {\overline {\bf{T}} }_{i,j}  -{\bf I}_{N_{\rm RF}}    \big)  \Big)\\
	&~~~+ T_{ij}^2 {\rm Tr}\Big(\big( {\bf{E}}_{i,j}^T {\overline {\bf{T}} }_{i,j}+ {\overline {\bf{T}} }_{i,j}^T  {\bf{E}}_{i,j}     \big)^2 \Big)\\
	&~~~+2 T_{ij}{\rm Tr}\Big(\big( {\bf{E}}_{i,j}^T {\overline {\bf{T}} }_{i,j}+ {\overline {\bf{T}} }_{i,j}^T  {\bf{E}}_{i,j}     \big) \big(  {\overline {\bf{T}} }_{i,j}^T  {\overline {\bf{T}} }_{i,j}  -{\bf I}_{N_{\rm RF}}    \big)  \Big)\\
	&={T_{ij}^4}+2T_{ij}^2 \big[{\overline {\bf{T}} }_{i,j}^T  {\overline {\bf{T}} }_{i,j}  -{\bf I}_{N_{\rm RF}}  \big]_{j,j} +2T_{ij}^2 \big[  {\overline {\bf{T}} }_{i,j} {\overline {\bf{T}} }_{i,j}^T   \big]_{i,i}\\
	&~~~+4T_{ij}    \big[ \overline {\bf{T}}  _{i,j} \big(  {\overline {\bf{T}} }_{i,j}^T  {\overline {\bf{T}} }_{i,j}  -{\bf I}_{N_{\rm RF}}    \big)   \big]_{i,j}     +{\rm const.}
	\end{aligned}
	\end{equation}
	where $ {\rm const.} $ is a constant term unrelated to $ T_{i,j} $. Thus, we attain 
	\begin{equation}
	\begin{aligned}
	& \frac{{\partial {\psi _2}(T_{i,j})}}{{\partial  T_{i,j}}}  \\
	& \quad ~~=4T_{i,j}^3+4T_{i,j}\left( \big[{\overline {\bf{T}} }_{i,j}^T  {\overline {\bf{T}} }_{i,j}  -{\bf I}_{N_{\rm RF}}  \big]_{j,j}+ \big[  {\overline {\bf{T}} }_{i,j} {\overline {\bf{T}} }_{i,j}^T   \big]_{i,i}  \right) \\
	&  \qquad~~~ + 4 \big[ \overline {\bf{T}}  _{i,j} \big(  {\overline {\bf{T}} }_{i,j}^T  {\overline {\bf{T}} }_{i,j}  -{\bf I}_{N_{\rm RF}}    \big)   \big]_{i,j} 
	\end{aligned}
	\end{equation}
}
	\ifCLASSOPTIONcaptionsoff
	\newpage
	\fi

	
	
	%
	
	\footnotesize
	\balance
	\bibliographystyle{IEEEtran}
	\bibliography{IEEEabrv,stan_ref}

\end{document}